\documentclass[preprint,prd,aps,showpacs,showkeys,nofootinbib]{revtex4}
\usepackage{graphicx}
\usepackage{dcolumn}
\usepackage{bm}
\usepackage{ulem}
\usepackage{color}
\usepackage[dvipsnames]{xcolor}
\usepackage{amssymb}
\usepackage{appendix}
\definecolor{light-gray}{gray}{0.78}
\definecolor{mid-gray}{gray}{0.55}
\definecolor{dark-gray}{gray}{0.32}

\begin{document}

\title{A 95 GeV Higgs Boson in the $U(1)_X$SSM}
\author{Song Gao$^{1,2,3}$, Shu-Min Zhao$^{1,2,3}$\footnote{zhaosm@hbu.edu.cn}, Shuang Di$^{1,2,3}$, Xing-Xing Dong$^{1,2,3,4}$, Tai-Fu Feng$^{1,2,3,5}$}
\affiliation{$^1$ Department of Physics, Hebei University, Baoding 071002, China}
\affiliation{$^2$ Hebei Key Laboratory of High-precision Computation and Application of Quantum Field Theory, Baoding, 071002, China}
\affiliation{$^3$ Hebei Research Center of the Basic Discipline for Computational Physics, Baoding, 071002, China}
\affiliation{$^4$ Departamento de F\'{\i}sica and CFTP, Instituto Superior T\'{e}cnico, Universidade de Lisboa, Av.Rovisco Pais 1,1049-001 Lisboa, Portugal}
\affiliation{$^5$ Department of Physics, Chongqing University, Chongqing 401331, China}
\date{\today}

\begin{abstract}
The CMS and ATLAS Collaborations have recently reported their findings based on the comprehensive run 2 dataset, detailing their searches for a light Higgs boson with a mass of approximately 95 GeV. We investigate the excesses observed in the $\gamma\gamma$ and $b{\bar b}$ data at approximately 95 GeV in the $U(1)_X$ extension of
 the minimal supersymmetric standard model ($U(1)_X$SSM). Additionally, it also mixes with the SM-like Higgs boson. Research indicates that, in this model, identifying the mixture of the singlet Higgs states as the lightest Higgs boson holds tremendous potential for explaining the excess observed at approximately 95 GeV. In our calculations, we maintain the masses of the lightest and next-to-lightest Higgs bosons at approximately 95 GeV and 125 GeV, respectively. The study finds that the theoretical predictions for the signal strengths $\mu(h_{95})_{\gamma\gamma}$ and $\mu(h_{95})_{b{\bar b}}$ in the $U(1)_X$SSM align well with the excesses observed by CMS.
\end{abstract}
\keywords{95GeV excesses, $U(1)_X$SSM, new physics}
\maketitle

\section{Introduction}
After the discovery of the Higgs boson by the ATLAS and CMS experiments at the Large Hadron Collider (LHC)\cite{FindH1,FindH2}, one of the main goals of the current LHC program is to investigate whether the detected Higgs boson is the only fundamental scalar particle as predicted by the Standard Model (SM), or whether it is part of a Beyond the Standard Model (BSM) theory with an extended Higgs sector and additional Higgs bosons. Therefore, one of the primary objectives of the current and future runs of the LHC is to search for additional Higgs bosons, which is crucial for exploring the fundamental physics mechanisms of electroweak symmetry breaking. These additional Higgs bosons can have masses either higher or lower than 125 GeV.

Searches for Higgs bosons with masses below 125 GeV have been conducted at the LEP\cite{LFP2,LFP3}, the Tevatron at Fermi National Accelerator Laboratory\cite{TFNAL}, and the LHC at CERN\cite{LHC1,LHC2,LHC3,LHC4}. Searches for $\gamma\gamma$ resonances at the LHC are particularly fascinating and hold great promise. This is evident from the fact that, owing to its relatively clean final state, this decay mode is one of the two discovery channels for the 125 GeV Higgs boson. CMS has conducted searches for scalar di-photon resonances at 8 TeV and 13 TeV. Results based on 8 TeV data, corresponding to an integrated luminosity of 19.7 fb$^{-1}$, and the first year of Run 2 data at 13 TeV, with an integrated luminosity of 35.9 fb$^{-1}$, reveal a local excess of 2.8$\sigma$ at 95.3 GeV\cite{LFP2,LFP3}, which is compatible with the latest ATLAS result\cite{ATLAS:2018xad}:
\begin{eqnarray}
&&\mu^{exp}_{\gamma\gamma}=\mu(\Phi_{95})^{ATLAS+CMS}_{\gamma\gamma}=0.24_{-0.08}^{+0.09}.
\end{eqnarray}

LEP reports a 2.3$\sigma$ local excess in the $b\bar b$ final state with mass around 95.4 GeV, the relevant
signal strength is\cite{LFP2}
\begin{eqnarray}
&&\mu(\Phi_{95})^{exp}_{b\bar b}=0.117\pm0.057.
\end{eqnarray}

Theoretically, there are numerous discussions on the excesses observed in NP. The analysis present in Refs.\cite{NEWNMSSM1,NEWNMSSM2,NEWNMSSM3,gamma,NMSSM,1,2,3,4,5,6,7,8,9,10} indicates that, in the Next-to-Minimal Supersymmetric Standard Model (NMSSM), the diphoton rate can be several times larger than the SM prediction for the same scalar mass. In the Two-Higgs Doublet Model with an additional real singlet (N2HDM), Refs.\cite{11,12,13,14,15} explore the possibilities of explaining the observed excesses. The authors of Ref.\cite{16} investigate whether a mixed radion-Higgs can serve as the 125 GeV boson, along with the presence of a light radion, to well account for the CMS $\gamma\gamma$ excess in the Higgs-radion mixing model.

In NMSSM, as discussed in Refs.\cite{NEWNMSSM1,NEWNMSSM2,NEWNMSSM3}, the additional Higgs boson is typically interpreted as a scalar particle dominated by the singlet field S. This reference thoroughly examines how the NMSSM explains the 95 GeV anomaly through the coupling of the S field with the Higgs doublets, emphasizing the critical role of the S field in modulating the Higgs mass and coupling properties. However, in our $U(1)_X$SSM model, numerical analysis reveals that the lightest Higgs boson is not primarily composed of the S field, but rather arises from a mixture of the $\eta$ and $\bar\eta$ fields. This conclusion is supported by the fact that $\sqrt{m_{{\phi}_{s}{\phi}_{s}}}$ reaches the order of $10^4$ GeV, which is significantly larger than the $10^2$ GeV scale of the lightest Higgs boson. This indicates that the contribution of the S field to the lightest Higgs boson is subordinate.
This interpretation not only enriches our understanding of the 95 GeV anomaly but also provides new avenues for future experimental verification.

 In the $U(1)_X$SSM, there is a new gauge boson$(A^{'X}_\mu)$ for the new gauge group $U(1)_X$.
Its neutralino supersymmetric partner is $\lambda_{\tilde{X}}$. There are five neutral CP-even Higgs component fields ($\hat{\eta},~\hat{\bar{\eta}},~\hat{S}$, $H_{u},~H_{d}$) in the model,
 and mix together, forming a $5\times 5$ mass-squared matrix. Consequently, the mass of the lightest CP-even Higgs particle can be improved at the tree level.
 In the $U(1)_X$SSM, the small hierarchy problem in the MSSM is alleviated through the added right-handed neutrinos, sneutrinos, and extra Higgs singlets. The $\mu$-problem existing in the MSSM is
  relieved after the spontaneous symmetry breaking of the $S$ field in vacuum through $\lambda_H\hat{S}\hat{H}_u\hat{H}_d$.
   Through the term $Y_\nu\hat{\nu} \hat{l} \hat{H}_u$, the right-handed neutrinos and
 left-handed neutrinos mix together\cite{U1X1,U1X2,U1X3,U1X4}, which makes light neutrinos to obtain tiny
 masses through the seesaw mechanism. Based on the characteristics of the newly introduced CP-even scalar in the $U(1)_X$SSM, we focus on investigating the excess phenomena of $\gamma\gamma$, and  $b\bar b$, at around 95 GeV in this work.

Due to the different composition of the lightest Higgs boson, our model exhibits novel features in its parameter space. For instance, parameters such as the mixing ratio between the $\eta$ and $\bar\eta$ fields and the scale of $U(1)_X$ symmetry breaking play crucial roles in explaining the 95 GeV anomaly. These parameters have not been the focus of attention in the NMSSM, and thus our work offers a new direction for the study of supersymmetric model parameter spaces.

 The outline of this paper is as follows: In Sec.II A, we introduce the fundamental elements of $U(1)_X$SSM, including its superpotential, general soft-breaking terms, and the tree-level mass-squared matrix for CP-even Higgs particle. In Sec.II B, we present in detail the formulas required for the corrections and the necessary mass matrices. In Sec.II C, we  provide the formulas for the two signals in detail. In Section III, we show the corresponding parameters and numerical analysis. In Section IV, we summarize this paper. Finally, some analytical results are given in the Appendix.
\section{Preliminary theories}

\subsection{The relevant content of $U(1)_X$SSM}

In the context of the $U(1)_X$SSM, the superpotential is presented as follows\cite{U1X3,U1X4,Gao2024,zsm1,zsm2}
\begin{eqnarray}
&&W=l_W\hat{S}+\mu\hat{H}_u\hat{H}_d+M_S\hat{S}\hat{S}-Y_d\hat{d}\hat{q}\hat{H}_d-Y_e\hat{e}\hat{l}\hat{H}_d+\lambda_H\hat{S}\hat{H}_u\hat{H}_d\nonumber
\\&&~~ ~ +\lambda_C\hat{S}\hat{\eta}\hat{\bar{\eta}}+\frac{\kappa}{3}\hat{S}\hat{S}\hat{S}+Y_u\hat{u}\hat{q}\hat{H}_u+Y_X\hat{\nu}\hat{\bar{\eta}}\hat{\nu}
+Y_\nu\hat{\nu}\hat{l}\hat{H}_u.
\end{eqnarray}

We use the notations of $v_{u}$, $v_{d}$, $v_{\eta}$, $v_{\bar{\eta}}$ and $v_{S}$ to signify the vacuum expectation values (VEVs) associated with the Higgs superfields $H_{u}$, $H_{d}$, $v_{\eta}$, $v_{\bar{\eta}}$, and $S$. Two angles are defined as $\tan\beta$ = $v_{u}$/$v_{d}$ and $\tan\beta_{\eta}$ = $v_{\bar{\eta}}$/$v_{\eta}$, respectively. Subsequently, we present the precise formulations of the two Higgs doublets and three Higgs singlets here
\begin{eqnarray}
&&\hspace{1cm}H_{u}=\left(\begin{array}{c}H_{u}^+\\{1\over\sqrt{2}}\Big(v_{u}+\phi_{u}+iP_{u}^0\Big)\end{array}\right),~~
H_{d}=\left(\begin{array}{c}{1\over\sqrt{2}}\Big(v_{d}+\phi_{d}+iP_{d}^0\Big)\\H_{d}^-\end{array}\right),
\nonumber\\&&\eta={1\over\sqrt{2}}\Big(v_{\eta}+\phi_{\eta}+iP_{\eta}^0\Big),~~~
\bar{\eta}={1\over\sqrt{2}}\Big(v_{\bar{\eta}}+\phi_{\bar{\eta}}+iP_{\bar{\eta}}^0\Big),~~
S={1\over\sqrt{2}}\Big(v_{S}+\phi_{S}+iP_{S}^0\Big).
\end{eqnarray}
And two angles are defined as $\tan\beta$ = $v_{u}$/$v_{d}$ and $\tan\beta_{\eta}$ = $v_{\bar{\eta}}$/$v_{\eta}$.

The soft SUSY breaking terms of $U(1)_X$SSM are as follows
\begin{eqnarray}
&&\mathcal{L}_{soft}=\mathcal{L}_{soft}^{MSSM}-B_SS^2-L_SS-\frac{T_\kappa}{3}S^3-T_{\lambda_C}S\eta\bar{\eta}
+\epsilon_{ij}T_{\lambda_H}SH_d^iH_u^j\nonumber\\&&\hspace{1cm}
-T_X^{IJ}\bar{\eta}\tilde{\nu}_R^{*I}\tilde{\nu}_R^{*J}
+\epsilon_{ij}T^{IJ}_{\nu}H_u^i\tilde{\nu}_R^{I*}\tilde{l}_j^J
-m_{\eta}^2|\eta|^2-m_{\bar{\eta}}^2|\bar{\eta}|^2-m_S^2S^2\nonumber\\&&\hspace{1cm}
-(m_{\tilde{\nu}_R}^2)^{IJ}\tilde{\nu}_R^{I*}\tilde{\nu}_R^{J}
-\frac{1}{2}\Big(M_{BL}\lambda^2_{\tilde{X}}+2M_{BB^\prime}\lambda_{\tilde{B}}\lambda_{\tilde{X}}\Big)+h.c.
\end{eqnarray}
$\mathcal{L}_{soft}^{MSSM}$ represents the soft breaking terms of MSSM.

\begin{table}
\caption{ The superfields in $U(1)_X$SSM}
\begin{tabular}{|c|c|c|c|c|c|c|c|c|c|c|c|}
\hline
Superfields & $\hspace{0.1cm}\hat{q}_i\hspace{0.1cm}$ & $\hat{u}^c_i$ & $\hspace{0.2cm}\hat{d}^c_i\hspace{0.2cm}$ & $\hat{l}_i$ & $\hspace{0.2cm}\hat{e}^c_i\hspace{0.2cm}$ & $\hat{\nu}_i$ & $\hspace{0.1cm}\hat{H}_u\hspace{0.1cm}$ & $\hat{H}_d$ & $\hspace{0.2cm}\hat{\eta}\hspace{0.2cm}$ & $\hspace{0.2cm}\hat{\bar{\eta}}\hspace{0.2cm}$ & $\hspace{0.2cm}\hat{S}\hspace{0.2cm}$ \\
\hline
$SU(3)_C$ & 3 & $\bar{3}$ & $\bar{3}$ & 1 & 1 & 1 & 1 & 1 & 1 & 1 & 1  \\
\hline
$SU(2)_L$ & 2 & 1 & 1 & 2 & 1 & 1 & 2 & 2 & 1 & 1 & 1  \\
\hline
$U(1)_Y$ & 1/6 & -2/3 & 1/3 & -1/2 & 1 & 0 & 1/2 & -1/2 & 0 & 0 & 0  \\
\hline
$U(1)_X$ & 0 & -1/2 & 1/2 & 0 & 1/2 & -1/2 & 1/2 & -1/2 & -1 & 1 & 0  \\
\hline
\end{tabular}
\label{JJ1}
\end{table}

In the context of $U(1)_X$SSM, the simultaneous existence of two Abelian groups, $U(1)_Y$ and $U(1)_X$, gives rise to a new effect that does not exist in MSSM: gauge kinetic mixing. This effect can also be induced through Renormalization Group Equations (RGEs), even if it is assumed as zero at $ M_{GUT}$.

$Y^Y$ denotes the $U(1)_Y$ charge, while $Y^X$ represents the $U(1)_X$ charge. One can write the covariant derivatives of the $U(1)_X$SSM in the following form
\begin{eqnarray}
&&D_\mu=\partial_\mu-i\left(\begin{array}{cc}Y^Y,&Y^X\end{array}\right)
\left(\begin{array}{cc}g_{Y},&g{'}_{{YX}}\\g{'}_{{XY}},&g{'}_{{X}}\end{array}\right)
\left(\begin{array}{c}A{'}_{\mu}^{Y} \\ A{'}_{\mu}^{X}\end{array}\right)\;,
\end{eqnarray}
where $A_{\mu}^{\prime Y}$ and $A_{\mu}^{\prime X}$ represent the gauge fields of $U(1)_Y$ and $U(1)_X$, respectively.

Under the condition that the two Abelian gauge groups remain unbroken, we employ the rotation matrix R to carry out a change of basis\cite{R}
\begin{eqnarray}
&&\left(\begin{array}{cc}g_{Y},&g{'}_{{YX}}\\g{'}_{{XY}},&g{'}_{{X}}\end{array}\right)
R^T=\left(\begin{array}{cc}g_{1},&g_{{YX}}\\0,&g_{{X}}\end{array}\right)~,~~~~
R\left(\begin{array}{c}A_{\mu}^{\prime Y} \\ A_{\mu}^{\prime X}\end{array}\right)
=\left(\begin{array}{c}A_{\mu}^{Y} \\ A_{\mu}^{X}\end{array}\right)\;,
\end{eqnarray}
In the end, the covariant derivatives of the $U(1)_X$SSM transform into
\begin{eqnarray}
&&D_\mu=\partial_\mu-i\left(\begin{array}{cc}Y^Y,&Y^X\end{array}\right)
\left(\begin{array}{cc}g_{1},&g_{{YX}}\\0,&g_{{X}}\end{array}\right)
\left(\begin{array}{c}A_{\mu}^{Y} \\ A_{\mu}^{X}\end{array}\right)\;.
\end{eqnarray}
At the tree level, three neutral gauge bosons $A_{\mu}^{\prime X}$, $A_{\mu}^{\prime Y}$ and $V_{\mu}^{\prime 3}$ mix with each other($V_{\mu}^{\prime 3}$ represents the neutral gauge field of $SU(2)_L$), with their mass matrix represented in the basis of ($A_{\mu}^{Y}$, $V_{\mu}^{3}$, $A_{\mu}^{X}$).
\begin{eqnarray}
\left(\begin{array}{ccc}\frac{1}{8}g_{1}^{2}v^{2},
&-\frac{1}{8}g_{1}g_{2}v^{2},&\frac{1}{8}g_{1}(g_{YX}+g_{X})v^{2}\\
-\frac{1}{8}g_{1}g_{2}v^{2},
&\frac{1}{8}g_{2}^2v^{2},&-\frac{1}{8}g_{2}(g_{X}+g_{YX})v^{2}\\
\frac{1}{8}g_{1}(g_{YX}+g_{X})v^{2},&-\frac{1}{8}g_{2}(g_{YX}+g_{X})v^{2},&\frac{1}{8}g_{X}^{2}\xi^{2}+\frac{1}{8}(g_{YX}+g_{X})^2v^{2}\end{array}\right),
\end{eqnarray}
with $v^2$ = $v_u^2$ + $v_d^2$ and $\xi^2$ = $v_\eta^2$ + $v_{\bar\eta}^2$. We deduce their mass eigenvalues as
\begin{eqnarray}
&&m_\gamma^2=0,\nonumber\\
&&m_{Z,{Z^{'}}}^2=\frac{1}{8}\Big((g_{1}^2+g_2^2+(g_{YX}+g_X)^2)v^2+4g_{X}^2\xi^2,\nonumber\\
&&\mp\sqrt{(g_{1}^2+g_{2}^2+(g_{YX}+g_X)^2)^2v^4+8((g_{YX}+g_X)^2-g_{1}^2-
g_{2}^2)g_{X}^2v^2\xi^2+16g_{X}^4\xi^4}\Big).
\end{eqnarray}

Assuming $\mu,~\lambda_H,~\lambda_C,~l_W,~ M_S,~B_\mu,~L_S,~T_{\kappa},~ T_{\lambda_C},~ T_{\lambda_H},~\kappa,~B_S$ are real parameters, we present the tree-level simplified Higgs potential as follows
\begin{eqnarray}
&&V_{Tree}=\frac{1}{2}g_X(g_X+g_{YX})(|H_d^0|^2-|H_u^0|^2)(|\eta|^2-|\bar{\eta}|^2)+\lambda_H^2|H_u^0H_d^0|^2+m^2_{S}|S|^2+l_W^2\nonumber\\&&~~ ~
+\frac{1}{8}\Big(g_1^2+g_2^2+(g_X+g_{YX})^2\Big)(|H_d^0|^2-|H_u^0|^2)^2+\frac{1}{2}g_X^2(|\eta|^2-|\bar{\eta}|^2)^2+\lambda_C^2|\eta\bar{\eta}|^2
\nonumber\\&&
~~ ~ +(\mu^2+\lambda_H^2|S|^2+2\mathrm{Re}[\mu\lambda_HS])(|H_d^0|^2+|H_u^0|^2)+\lambda_C^2|S|^2(|\eta|^2+|\bar{\eta}|^2)+m^2_\eta|\eta|^2
\nonumber\\&& ~~ ~ +2\mathrm{Re}[(l_W+2 M_SS^*)(\lambda_C\eta\bar{\eta}-\lambda_HH_u^0H_d^0
+\kappa S^2)]+4M_S^2|S|^2+\kappa^2|S|^4+m^2_{\bar{\eta}}|\bar{\eta}|^2\nonumber\\&& ~~ ~ +
2\mathrm{Re}[\lambda_C\kappa\eta^*\bar{\eta}^*S^2+2l_WM_SS
-\lambda_C\lambda_H\eta^*\bar{\eta}^*H_u^0H_d^0]+
m^2_{H_u^0}|H_u|^2+m^2_{H_d}|H_d|^2\nonumber\\&& ~~ ~
+ 2\mathrm{Re}\Big[L_S S- H_d^0H_u^0\Big(B_\mu+\lambda_H\kappa  (S^2)^*+T_{\lambda_H} S\Big)
+\frac{1}{3}T_kS^3+ T_{\lambda_C}\eta\bar{\eta}S+B_S S^2\Big].\label{Vtree}
\end{eqnarray}

In Ref.\cite{U1X3}, the corresponding tadpole equations at the tree level are also derived. Additionally, the tree-level mass-squared matrix for the CP-even Higgs $({\phi}_{d}, {\phi}_{u}, {\phi}_{\eta}, {\phi}_{\overline{\eta}}, {\phi}_{s})$ is also presented
\begin{eqnarray}
M^2_{h,tree} = \left(
\begin{array}{ccccc}
m_{{\phi}_{d}{\phi}_{d}} &m_{{\phi}_{u}{\phi}_{d}} &m_{{\phi}_{\eta}{\phi}_{d}} &m_{{\phi}_{\bar{\eta}}{\phi}_{d}} &m_{{\phi}_{s}{\phi}_{d}}\\
m_{{\phi}_{d}{\phi}_{u}} &m_{{\phi}_{u}{\phi}_{u}} &m_{{\phi}_{\eta}{\phi}_{u}} &m_{{\phi}_{\bar{\eta}}{\phi}_{u}} &m_{{\phi}_{s}{\phi}_{u}}\\
m_{{\phi}_{d}{\phi}_{\eta}} &m_{{\phi}_{u}{\phi}_{\eta}} &m_{{\phi}_{\eta}{\phi}_{\eta}} &m_{{\phi}_{\bar{\eta}}{\phi}_{\eta}} &m_{{\phi}_{s}{\phi}_{\eta}}\\
m_{{\phi}_{d}{\phi}_{\bar{\eta}}} &m_{{\phi}_{u}{\phi}_{\bar{\eta}}} &m_{{\phi}_{\eta}{\phi}_{\bar{\eta}}} &m_{{\phi}_{\bar{\eta}}{\phi}_{\bar{\eta}}} &m_{{\phi}_{s}{\phi}_{\bar{\eta}}}\\
m_{{\phi}_{d}{\phi}_{s}} &m_{{\phi}_{u}{\phi}_{s}} &m_{{\phi}_{\eta}{\phi}_{s}} &m_{{\phi}_{\bar{\eta}}{\phi}_{s}} &m_{{\phi}_{s}{\phi}_{s}}\end{array}
\right),\label{Rsneu}
 \end{eqnarray}
\begin{eqnarray}
&&m_{\phi_{d}\phi_{d}}= m_{H_d}^2+  \mu^2
 +\frac{1}{8} \Big( [g_{1}^{2}+(g_{X}+g_{YX})^{2}+g_2^2] (3 v_{d}^{2}  - v_{u}^{2})\nonumber \\
&&\hspace{1.5cm}+2 (g_{Y X} g_{X}+g_X^2) ( v_{\eta}^{2}- v_{\bar{\eta}}^{2})\Big)+ \sqrt{2} v_S \mu {\lambda}_{H}+\frac{1}{2} (v_{u}^{2} + v_S^{2}){\lambda}_{H}^2,
\\&&m_{\phi_{d}\phi_{u}} = -\frac{1}{4} \Big(g_{2}^{2} + (g_{Y X} + g_{X})^{2} + g_1^{2}\Big)v_d v_u
 + {\lambda}_{H}^2 v_d v_u - {\lambda}_{H} l_W \nonumber \\
&&\hspace{1.5cm}-\frac{1}{2}{\lambda}_{H} (v_{\eta} v_{\bar{\eta}} {\lambda}_{C}  + v_S^{2} \kappa )
 - B_{\mu}- \sqrt{2} v_S (\frac{1}{2}T_{{\lambda}_{H}}  + M_S {\lambda}_{H} ),
\\ &&m_{\phi_{u}\phi_{u}} = m_{H_u}^2+ \mu^2+\frac{1}{8} \Big( [g_{1}^{2}+(g_{X}+g_{YX})^{2}+g_2^2] (3 v_{u}^{2}  - v_{d}^{2})\nonumber \\
&&\hspace{1.5cm}+2 (g_{Y X} g_{X}+g_X^2) ( v_{\bar{\eta}}^{2}-v_{\eta}^{2})\Big)
 +  \sqrt{2} v_S\mu {\lambda}_{H}   + \frac{1}{2}(v_{d}^{2} + v_S^{2}){\lambda}_{H}^2,
\\&&m_{\phi_{d}\phi_{\eta}} = \frac{1}{2}g_{X} (g_{Y X} + g_{X})v_d v_{\eta}
  -\frac{1}{2} v_u v_{\bar{\eta}} {\lambda}_{H} {\lambda}_{C} ,
\\&&m_{\phi_{u}\phi_{\eta}} = -\frac{1}{2}g_{X} (g_{Y X} + g_{X})v_u v_{\eta}
-\frac{1}{2} v_d v_{\bar{\eta}} {\lambda}_{H} {\lambda}_{C},
\\&&m_{\phi_{\eta}\phi_{\eta}} = m_{\eta}^2 +\frac{1}{4} \Big((g_{Y X} g_{X}+g_X^2) ( v_{d}^{2}
- v_{u}^{2})+2g_{X}^{2}
( 3 v_{\eta}^{2}-v_{\bar{\eta}}^{2})\Big)+\frac{{\lambda}_{C}^2}{2} (v_{\bar{\eta}}^{2} + v_S^{2}),
\\&&m_{\phi_{d}\phi_{\bar{\eta}}} = -\frac{1}{2}g_{X} (g_{Y X} + g_{X})v_d v_{\bar{\eta}}
  -\frac{1}{2} v_u v_{\eta} {\lambda}_{H} {\lambda}_{C},
\\&&m_{\phi_{u}\phi_{\bar{\eta}}} = \frac{1}{2}g_{X} (g_{Y X} + g_{X})v_u v_{\bar{\eta}}
 -\frac{1}{2} v_d v_{\eta} {\lambda}_{H} {\lambda}_{C},
\\&&m_{\phi_{\eta}\phi_{\bar{\eta}}} = ({\lambda}_{C}^2- g_{X}^{2})v_{\eta} v_{\bar{\eta}}+\frac{{\lambda}_{C}}{2}(2 l_W  - {\lambda}_{H} v_d v_u )
+ \frac{v_S }{\sqrt{2}} (2 M_S {\lambda}_{C} + T_{{\lambda}_{C}}) + \frac{v_S^{2}}{2} {\lambda}_{C} \kappa,
\\&&m_{\phi_{\bar{\eta}}\phi_{\bar{\eta}}} = m_{\bar{\eta}}^2+\frac{1}{4} \Big((g_{Y X} g_{X}+g_X^2)
 ( v_{u}^{2}- v_{d}^{2})+2g_{X}^{2}( 3 v_{\bar{\eta}}^{2}-v_{\eta}^{2})\Big)+\frac{{\lambda}_{C}^2 }{2} \Big(v_{\eta}^{2} + v_S^{2}\Big),
\\&&m_{\phi_{d}{\phi}_{s}} = \Big({\lambda}_{H} v_d v_S  + \sqrt{2} v_d \mu  -  v_u ( \kappa v_S
 + \sqrt{2} M_S )\Big){\lambda}_{H} - \frac{1}{\sqrt{2}}v_u T_{{\lambda}_{H}},
\\&&m_{\phi_{u}{\phi}_{s}} =  \Big( {\lambda}_{H} v_u v_S  + \sqrt{2} v_u \mu
-v_d (\kappa v_S  + \sqrt{2} M_S )\Big){\lambda}_{H}
- \frac{1}{\sqrt{2}}  v_dT_{{\lambda}_{H}},
\\&&m_{\phi_{\eta}{\phi}_{s}} = \Big( {\lambda}_{C} v_{\eta} v_S  + v_{\bar{\eta}} (\kappa v_S
 + \sqrt{2} M_S )\Big){\lambda}_{C}  +\frac{1}{\sqrt{2}}v_{\bar{\eta}} T_{{\lambda}_{C}},
\\&&m_{\phi_{\bar{\eta}}{\phi}_{s}} =\Big( {\lambda}_{C} v_{\bar{\eta}} v_S  + v_{\eta}(\kappa v_S
 + \sqrt{2} M_S )\Big){\lambda}_{C} + \frac{1}{\sqrt{2}}v_{\eta}
 T_{{\lambda}_{C}},
\\&&m_{{\phi}_{s}{\phi}_{s}} = m^2_{S}+ \Big(2 l_W  + 3v_S (\kappa v_S  + 2\sqrt{2} M_S )
+ {\lambda}_{C} v_{\eta} v_{\bar{\eta}}  - {\lambda}_{H} v_d v_u \Big)\kappa+2 {B_{S}}\\
&&\hspace{1.5cm} +\frac{1}{2}{\lambda}_{C}^2 \xi^2+\frac{1}{2}{\lambda}_{H}^2 v^{2}
  + 4 M_S^2   + \sqrt{2} v_S T_{\kappa}.\nonumber
\end{eqnarray}
\subsection{Higgs mass correction}
The effective potential at the one-loop level can be expressed in the form given below
\begin{eqnarray}
&&V_{Total}=V_{Tree}+\Delta V.
\end{eqnarray}

Here, $V_{Tree}$ represents the effective potential at the tree level, and $\Delta{V}$ is the potential from the one-loop correction. With the adoption of Dimensional Reduction renormalization scheme, the effective Higgs potential up to the one-loop correction is presented in the Landau gauge, and the specific form of $\Delta V$ is given as\cite{GS1,GS2}
\begin{eqnarray}
&&\Delta V=\sum_i\frac{n_i}{64\pi^2}m_i^4(\phi_d,\phi_u,\phi_\eta, \phi_{\bar{\eta}},\phi_s)\Big(
\log\frac{m_i^2(\phi_d,\phi_u,\phi_\eta, \phi_{\bar{\eta}},\phi_s)}{Q^2}-\frac{3}{2}\Big).
\end{eqnarray}

We set the renormalization scale Q to be in the TeV order, and denote the degrees of freedom for each mass eigenstate by $n_i$ (-12 for quarks, -4 for leptons and charginos, -2 for neutralinos and neutrinos,
6 for squarks, 2 for sleptons and charged Higgs, 3 and 6 for $Z(Z^\prime)$ and $W$ bosons,
 1 for sneutrinos and the neutral Higgs scalars). The specific potential for the one-loop correction is as follows
\begin{eqnarray}
&&\Delta{V}=V_t+V_b+V_\tau+V_{\tilde{u}_i}+V_{\tilde{d}_i}+V_{\tilde{e}_i}+V_{\tilde{\nu}}+V_{W}+V_{Z,Z'}+V_{\tilde{\chi}^0}+V_{H^-}+V_{\tilde{\chi}^\pm}.
\end{eqnarray}
Here, $V_t$, $V_b$ and $V_\tau$ represent the one-loop effective potential corrections from $t$, $b$ and $\tau$.
$V_{\tilde{u}_i}$ and $V_{\tilde{d}_i}$ represent
the corrections from scalar up type quarks and scalar down type quarks. $V_{\tilde{e}_i}$ represents the corrections from slepton.
$V_{\tilde{\nu}}$ represent corrections from CP-even sneutrino and CP-odd sneutrino. $V_{W}$, $V_{Z,Z'}$, $V_{\tilde{\chi}^0}$, $V_{H^-}$ and $V_{\tilde{\chi}^\pm}$ represent the corrections from $W$ boson, $Z(Z')$ boson, neutralinos, charged Higgs and charginos, respectively.

The mass matrices are required, and we gather the mass matrices for the CP-even sneutrino, CP-odd sneutrino, slepton, squark, chargino, charged Higgs and neutralino.

The mass matrix for CP-even sneutrino $({\phi}_{l}, {\phi}_{r})$ reads
\begin{eqnarray}
M^2_{\tilde{\nu}^R} = \left(
\begin{array}{cc}
m_{{\phi}_{l}{\phi}_{l}} &m^T_{{\phi}_{r}{\phi}_{l}}\\
m_{{\phi}_{l}{\phi}_{r}} &m_{{\phi}_{r}{\phi}_{r}}\end{array}
\right),\label{Rsneu}
 \end{eqnarray}
\begin{eqnarray}
&&m_{{\phi}_{l}{\phi}_{l}}= \frac{1}{8} \Big((g_{1}^{2} + g_{Y X}^{2} + g_{2}^{2}+ g_{Y X} g_{X})( v_{d}^{2}- v_{u}^{2})
+  2g_{Y X} g_{X}(v_{\eta}^{2}- v_{\bar{\eta}}^{2})\Big)
+\frac{ v_{u}^{2}}{2}{Y_{\nu}^2}  + m_{\tilde{L}}^2\nonumber,
 \\&&m_{{\phi}_{l}{\phi}_{r}} = \frac{1}{\sqrt{2} } v_uT_\nu  +  v_u v_{\bar{\eta}} {Y_X  Y_\nu}
  - \frac{1}{2}v_d ({\lambda}_{H}v_S  + \sqrt{2} \mu )Y_\nu,\nonumber\\&&
m_{{\phi}_{r}{\phi}_{r}}= \frac{1}{8} \Big((g_{Y X} g_{X}+g_{X}^{2})(v_{d}^{2}- v_{u}^{2})
+2g_{X}^{2}(v_{\eta}^{2}- v_{\bar{\eta}}^{2})\Big) + v_{\eta} v_S Y_X {\lambda}_{C}\nonumber \\&&\hspace{1.8cm}
 +m_{\tilde{\nu}}^2 + \frac{1}{2} v_{u}^{2}|Y_\nu|^2+  v_{\bar{\eta}} (2 v_{\bar{\eta}}Y_X ^2  + \sqrt{2} T_X).
\end{eqnarray}
To obtain the masses of sneutrinos, the rotation matrix $Z^R$ is used to diagonalize $M^2_{\tilde{\nu}^R}$.

We also deduce the mass matrix for CP-odd sneutrino $({\sigma}_{l}, {\sigma}_{r})$
\begin{eqnarray}
M^2_{\tilde{\nu}^I} = \left(
\begin{array}{cc}
m_{{\sigma}_{l}{\sigma}_{l}} &m^T_{{\sigma}_{r}{\sigma}_{l}}\\
m_{{\sigma}_{l}{\sigma}_{r}} &m_{{\sigma}_{r}{\sigma}_{r}}\end{array}
\right),
 \end{eqnarray}
\begin{eqnarray}
&&m_{{\sigma}_{l}{\sigma}_{l}}= \frac{1}{8} \Big((g_{1}^{2} + g_{Y X}^{2} + g_{2}^{2}+  g_{Y X} g_{X})( v_{d}^{2}- v_{u}^{2})
+  2g_{Y X} g_{X}(v_{\eta}^{2}-v_{\bar{\eta}}^{2})\Big)
+\frac{ v_{u}^{2}}{2}Y_{\nu}^2  + m_{\tilde{L}}^2\nonumber,
 \\&&m_{{\sigma}_{l}{\sigma}_{r}} = \frac{1}{\sqrt{2} } v_uT_\nu -  v_u v_{\bar{\eta}} {Y_X  Y_\nu}
  - \frac{1}{2}v_d ({\lambda}_{H}v_S  + \sqrt{2} \mu )Y_\nu,\nonumber\\&&
m_{{\sigma}_{r}{\sigma}_{r}}= \frac{1}{8} \Big((g_{Y X} g_{X}+g_{X}^{2})(v_{d}^{2}- v_{u}^{2})
+2g_{X}^{2}(v_{\eta}^{2}- v_{\bar{\eta}}^{2})\Big)- v_{\eta} v_S Y_X {\lambda}_{C}\nonumber \\&&\hspace{1.8cm}
+m_{\tilde{\nu}}^2 + \frac{1}{2} v_{u}^{2}|Y_\nu|^2+  v_{\bar{\eta}} (2 v_{\bar{\eta}}Y_X  Y_X  - \sqrt{2} T_X).
\end{eqnarray}
 We use $Z^I$ to diagonalize the mass squared matrix of the sneutrino $M^2_{\tilde{\nu}^I}$.

In the basis $(\tilde{e}_L, \tilde{e}_R)$, the mass matrix for slepton is shown and diagonalized by $Z^E$ through the
formula $Z^E m^2_{\tilde{e}} Z^{E,\dagger} = m^{diag}_{2,\tilde{e}}$,
\begin{equation}
m^2_{\tilde{e}} = \left(
\begin{array}{cc}
m_{\tilde{e}_L\tilde{e}_L^*} &\frac{1}{2} \Big(\sqrt{2} v_d T_{e}^{\dagger}  - v_u ({\lambda}_{H} v_S  + \sqrt{2} \mu )Y_{e}^{\dagger} \Big)\\
\frac{1}{2} \Big(\sqrt{2} v_d T_e  - v_u Y_e (\sqrt{2} \mu  + v_S {\lambda}_{H} )\Big) &m_{\tilde{e}_R\tilde{e}_R^*}\end{array}
\right),
 \end{equation}
\begin{eqnarray}
&&m_{\tilde{e}_L\tilde{e}_L^*} = m_{\tilde{L}}^2+\frac{1}{8} \Big((g_{1}^{2} + g_{Y X}^{2}
+ g_{Y X} g_{X} -g_2^2)(v_{d}^{2}- v_{u}^{2})+ 2 g_{Y X} g_{X}( v_{\eta}^{2}- v_{\bar{\eta}}^{2}
)
\Big)+\frac{v_{d}^{2}}{2} {Y_{e}^2} ,\nonumber\\&&
m_{\tilde{e}_R\tilde{e}_R^*} = m_{\tilde{E}}^2-\frac{1}{8}  \Big([2(g_{1}^{2} + g_{Y X}^{2})+3g_{Y X} g_{X}+g_{X}^{2}]
( v_{d}^{2}- v_{u}^{2})\nonumber\\&&\hspace{1.7cm}+(4g_{Y X} g_{X}+2g_{X}^{2})(v_{\eta}^{2}- v_{\bar{\eta}}^{2})
\Big)+\frac{1}{2} v_{d}^{2} {  Y_{e}^2}.
\end{eqnarray}

The  squared mass matrix for down type squark is shown
 in the basis $\left(\tilde{d}^0_{L}, \tilde{d}^0_{R}\right)$
\begin{equation}
M^2_{\tilde{D}} = \left(
\begin{array}{cc}
m_{\tilde{d}_L^0\tilde{d}_L^{0,*}} &m^\dagger_{\tilde{d}_R^0\tilde{d}_L^{0,*}}\\
m_{\tilde{d}_L^0\tilde{d}_R^{0,*}} &m_{\tilde{d}_R^0\tilde{d}_R^{0,*}}\end{array}
\right),
\end{equation}

\begin{eqnarray}
&&m_{\tilde{d}_L^0\tilde{d}_L^{0,*}} = \frac{1}{24}
 \Big( (3 g_{2}^{2}  + g_{1}^{2} + g_{Y X}^{2}+  g_{Y X} g_{X}) ( v_{u}^{2}- v_{d}^{2}  ) +  2g_{Y X} g_{X}
 ( v_{\bar{\eta}}^{2}  - v_{\eta}^{2})\Big)+m_{\tilde{Q}}^2  +\frac{ v_{d}^{2}}{2} {Y_{d}^2},\nonumber\\
&&m_{\tilde{d}_L^0\tilde{d}_R^{0,*}} = -\frac{1}{2}  \Big(\sqrt{2}  (- v_d T_d  + v_u Y_d \mu ) + v_u v_S Y_d {\lambda}_{H} \Big),\nonumber\\
&&m_{\tilde{d}_R^0\tilde{d}_R^{0,*}} = \frac{1}{24}   \Big(  (2 g_{1}^{2} + 2g_{Y X}^{2}+ 5g_{Y X} g_{X}+3g_{X}^{2})
 ( v_{u}^{2}  - v_{d}^{2})+ 2(2g_{Y X} g_{X}+3 g_{X}^{2} )( v_{\bar{\eta}}^{2}  - v_{\eta}^{2})\Big)\nonumber\\&&\hspace{1.8cm} +m_{\tilde{D}}^2  +\frac{ v_{d}^{2}}{2} {Y_{d}^2}.
\end{eqnarray}

In the basis $\left(\tilde{u}^0_{L}, \tilde{u}^0_{R}\right)$, the squared mass matrix for up type squark is
\begin{equation}
M^2_{\tilde{U}} = \left(
\begin{array}{cc}
m_{\tilde{u}_L^0\tilde{u}_L^{0,*}} &m^\dagger_{\tilde{u}_R^0\tilde{u}_L^{0,*}}\\
m_{\tilde{u}_L^0\tilde{u}_R^{0,*}} &m_{\tilde{u}_R^0\tilde{u}_R^{0,*}}\end{array}
\right),
\end{equation}

\begin{eqnarray}
	&&m_{\tilde{u}_L^0\tilde{u}_L^{0,*}} = \frac{1}{24}  \Big( (g_{1}^{2} -3 g_{2}^{2}+ g_{Y X}^{2}+   g_{Y X} g_{X}) ( v_{u}^{2}- v_{d}^{2}  )
 +   g_{Y X} g_{X}  (2 v_{\bar{\eta}}^{2}  -2 v_{\eta}^{2} )\Big)
+  m_{\tilde{Q}}^2  +\frac{ v_{u}^{2}}{2} {Y_{u}^2},\nonumber\\
	&&m_{\tilde{u}_L^0\tilde{u}_R^{0,*}} = -\frac{1}{2}   \Big(\sqrt{2}  (v_d Y_u \mu  - v_u T_u ) + v_d v_S Y_u {\lambda}_{H} \Big),\nonumber\\
	&&m_{\tilde{u}_R^0\tilde{u}_R^{0,*}} = \frac{1}{24}   \Big( (4g_{1}^{2} + 4g_{Y X}^{2}+  7g_{Y X} g_{X}+3  g_{X}^{2})
( v_{d}^{2}- v_{u}^{2})+  2(4g_{Y X} g_{X} +3  g_{X}^{2}) (  v_{\eta}^{2}-v_{\bar{\eta}}^{2}  )\Big) \nonumber \\
	&&\hspace{1.8cm}+  m_{\tilde{U}}^2  +\frac{ v_{u}^{2}}{2} {Y_{u}^2}.\nonumber
\end{eqnarray}

In the basis $(\lambda_{\tilde{B}}, \tilde{W}^0, \tilde{H}_d^0, \tilde{H}_u^0,
\lambda_{\tilde{X}}, \tilde{\eta}, \tilde{\bar{\eta}}, \tilde{s}) $,
the mass matrix for neutralino is
\begin{equation}
m_{\tilde{\chi}^0} = \left(
\begin{array}{cccccccc}
M_1 &0 &-\frac{g_1}{2}v_d &\frac{g_1}{2}v_u &{M}_{B B'} &0  &0  &0\\
0 &M_2 &\frac{1}{2} g_2 v_d  &-\frac{1}{2} g_2 v_u  &0 &0 &0 &0\\
-\frac{g_1}{2}v_d &\frac{1}{2} g_2 v_d  &0
&m_{\tilde{H}_u^0\tilde{H}_d^0} &m_{\lambda_{\tilde{X}}\tilde{H}_d^0} &0 &0 & - \frac{{\lambda}_{H} v_u}{\sqrt{2}}\\
\frac{g_1}{2}v_u &-\frac{1}{2} g_2 v_u  &m_{\tilde{H}_d^0\tilde{H}_u^0} &0 &m_{\lambda_{\tilde{X}}\tilde{H}_u^0} &0 &0 &- \frac{{\lambda}_{H} v_d}{\sqrt{2}}\\
{M}_{B B'} &0 &m_{\tilde{H}_d^0\lambda_{\tilde{X}}} &m_{\tilde{H}_u^0\lambda_{\tilde{X}}} &{M}_{BL} &- g_{X} v_{\eta}  &g_{X} v_{\bar{\eta}}  &0\\
0  &0 &0 &0 &- g_{X} v_{\eta}  &0 &\frac{1}{\sqrt{2}} {\lambda}_{C} v_S  &\frac{1}{\sqrt{2}} {\lambda}_{C} v_{\bar{\eta}} \\
0  &0 &0 &0 &g_{X} v_{\bar{\eta}}  &\frac{1}{\sqrt{2}} {\lambda}_{C} v_S  &0 &\frac{1}{\sqrt{2}} {\lambda}_{C} v_{\eta} \\
0 &0 & - \frac{{\lambda}_{H} v_u}{\sqrt{2}} &- \frac{{\lambda}_{H} v_d}{\sqrt{2}} &0 &\frac{1}{\sqrt{2}} {\lambda}_{C} v_{\bar{\eta}}
 &\frac{1}{\sqrt{2}} {\lambda}_{C} v_{\eta}  &m_{\tilde{s}\tilde{s}}\end{array}
\right),\label{neutralino}
 \end{equation}

\begin{eqnarray}
&& m_{\tilde{H}_d^0\tilde{H}_u^0} = - \frac{1}{\sqrt{2}} {\lambda}_{H} v_S  - \mu ,~~~~~~~
m_{\tilde{H}_d^0\lambda_{\tilde{X}}} = -\frac{1}{2} (g_{Y X} + g_{X})v_d, \nonumber\\&&
m_{\tilde{H}_u^0\lambda_{\tilde{X}}} = \frac{1}{2} (g_{Y X} + g_{X})v_u
 ,~~~~~~~~~~~~
m_{\tilde{s}\tilde{s}} = 2 M_S  + \sqrt{2} \kappa v_S.\label{neutralino1}
\end{eqnarray}
This matrix is diagonalized by $Z^N$
\begin{equation}
Z^{N*} m_{\tilde{\chi}^0} Z^{N{\dagger}} = m^{diag}_{\tilde{\chi}^0}.
\end{equation}

The mass matrix for charged Higgs in the basis: ($H_d^{-}$,$H_u^{+,*}$), ($H_d^{-,*}$,$H_u^{+}$)
\begin{eqnarray}
M_{H^-}^2=
\left({\begin{array}{*{20}{c}}
m_{{H_d^{-}}H_d^{-,*}} & m_{H_u^{+,*}H_d^{-,*}}^{*} \\
m_{H_d^{-}H_u^{+}} & m_{H_u^{+,*}H_u^{+}} \\
\end{array}}
\right),
\end{eqnarray}
\begin{eqnarray}
&&m_{{H_d^{-}}H_d^{-,*}}=\frac{1}{8}\Big((g_2^2+g_X^2)v_d^2+(g_2^2-g_X^2)v_u^2+(g_1^2+g_{YX}^2)(v_d^2-v_u^2)-2g_X^2v_{\bar{\eta}}^2
\nonumber\\&&\hspace{1.8cm}+2[g_{YX}g_X(v_d^2+v_\eta^2-v_{\bar{\eta}}^2-v_u^2)+g_X^2v_\eta^2]\Big)\nonumber\\&&\hspace{1.8cm}
+\Big(\mid\mu\mid^2+\sqrt{2}v_S\Re(\mu\lambda_H^*)+\frac{1}{2}v_S^2\mid\lambda_H\mid^2 \Big),\\
&&m_{H_d^{-}H_u^{+}}=\frac{1}{2}\Big(2(\lambda_Hl_W^*+B_\mu)+\lambda_H(2\sqrt{2}v_S M_S^*-v_dv_u\lambda_H^*+v_\eta v_{\bar{\eta}}\lambda_C^*\nonumber\\&&\hspace{1.8cm}+\sqrt{2}v_S T_{\lambda_H})\Big)
+\frac{1}{4}g_2^2v_dv_u,\\
&&m_{H_u^{+,*}H_u^{+}}=\frac{1}{8}\Big((g_2^2-g_X^2)v_d^2+(g_2^2+g_X^2)v_u^2+(g_1^2+g_{YX}^2)(v_u^2-v_d^2)-2g_X^2v_\eta^2
\nonumber\\&&\hspace{1.8cm}+2[g_{YX}g_X(v_u^2+v_{\bar{\eta}}^2-v_d^2-v_\eta^2)+g_X^2v_{\bar{\eta}}^2]\Big)
\nonumber\\&&\hspace{1.8cm}+\frac{1}{2}\Big(2\mid\mu\mid^2+2\sqrt{2}v_S\Re(\mu\lambda_H^*)+v_S^2\mid\lambda_H\mid^2 \Big).
\end{eqnarray}

In the basis $ \left(\tilde{W}^-, \tilde{H}_d^-\right), \left(\tilde{W}^+, \tilde{H}_u^+\right)$, the definition of the mass matrix for charginos is given by
\begin{equation}
M_{\tilde{\chi}^\pm} = \left(
\begin{array}{cc}
M_2 &\frac{1}{\sqrt{2}} g_2 v_u \\
\frac{1}{\sqrt{2}} g_2 v_d  &\frac{1}{\sqrt{2}} {\lambda}_{H} v_S  + \mu\end{array}
\right).
\label{mxzf}
\end{equation}
This matrix is diagonalized by $U$ and $V$
\begin{eqnarray}
U^*M_{\tilde{\chi}^\pm}V^\dagger=M_{\tilde{\chi}^\pm}^{diag}.
\end{eqnarray}

Because the stability conditions will be used in the subsequent calculations, here we present the stability conditions at the one-loop level in the $U(1)_X$SSM
\begin{eqnarray}
&&\left \langle \frac{\partial V_{Total}}{\partial \phi_u} \right \rangle=
\left \langle \frac{\partial V_{Total}}{\partial \phi_d} \right \rangle=
\left \langle \frac{\partial V_{Total}}{\partial \phi_\eta} \right \rangle=
\left \langle \frac{\partial V_{Total}}{\partial \phi_{\bar{\eta}}} \right \rangle=
\left \langle \frac{\partial V_{Total}}{\partial \phi_s} \right \rangle=0.
\end{eqnarray}

The corresponding analysis results are quite cumbersome, so we adopt numerical methods to solve the equations. To save space in the article, we place the analysis results for a portion of the particles in the appendix.

The one-loop contributions in the effective potential $V_{Total}$ modify the mass-squared matrix of the CP-even Higgs
\begin{eqnarray}
M^2_{Total,h}=M^2_{Tree,h}+\Delta M^2_h.
\end{eqnarray}

The components of the adjusted mass-squared matrix $M^2_{Total,h,ij}$ can be inferred from the one-loop effective potential $V_{Total}$ using the following equation
\begin{eqnarray}
 M^2_{Total,h,ij}=\Big\langle\frac{\partial^2V_{Total}}{\partial \phi_i \partial \phi_j}\Big|_{\phi_i,\phi_j=v_d,v_u,v_\eta, v_{\bar{\eta}},v_S}\Big\rangle.
 \end{eqnarray}

The smallest eigenvalue of $M^2_{Total,h}$ should be the square of $m_{h_1}$ (95 GeV), and the second smallest eigenvalue should be the square of $m_{h_2}$ (125 GeV).
\subsection{$\gamma\gamma$ and $b\bar b$ signals}
As defined in the introduction sector,  the signal strengths for $\gamma\gamma$ and $b\bar b$ events are
\begin{eqnarray}
&&\mu^{m_{h_1}}_{\gamma\gamma}=\frac{\sigma(gg\rightarrow h^{\rm NP}_{95}){\rm BR}(h^{\rm NP}_{95}\rightarrow \gamma\gamma)}{\sigma(gg\rightarrow h^{\rm SM}_{95}){\rm BR}(h^{\rm SM}_{95}\rightarrow \gamma\gamma)},\nonumber\\
&&\mu^{m_{h_1}}_{b\bar b}=\frac{\sigma(Z^*\rightarrow Zh^{\rm NP}_{95}){\rm BR}(h^{\rm NP}_{95}\rightarrow b\bar b)}{\sigma(Z^*\rightarrow Zh^{\rm SM}_{95}){\rm BR}(h^{\rm SM}_{95}\rightarrow b\bar b)},
\end{eqnarray}
here\cite{SBKD}
\begin{eqnarray}
&&\Gamma^{\rm SM}_{\rm tot,95}\approx 0.00259\;{\rm GeV},~~ {\rm BR}(h^{\rm SM}_{95}\rightarrow \gamma\gamma)\approx 1.4\times10^{-3},\nonumber\\&&{\rm BR}(h^{\rm SM}_{95}\rightarrow b\bar b)\approx 0.801.
\end{eqnarray}
The contributions from NP can be written as\cite{Ge2024}
\begin{eqnarray}
&&\Gamma^{\rm NP}_{{\rm tot},95}\approx C_{h_{95}dd}^2\Gamma^{\rm SM}_{b\bar b,95}+C_{h_{95}ee}^2\Gamma^{\rm SM}_{\tau\bar \tau,95}+C_{h_{95}uu}^2(\Gamma^{\rm SM}_{c\bar c,95}+\Gamma^{\rm SM}_{gg,95}),\nonumber\\
&&{\rm BR}(h^{\rm NP}_{95}\rightarrow \gamma\gamma)\approx C_{h_{95}uu}^2 {\rm BR}(h^{\rm SM}_{95}\rightarrow \gamma\gamma)\frac{\Gamma^{\rm SM}_{\rm tot,95}}{\Gamma^{\rm NP}_{\rm tot,95}},\nonumber\\
&&{\rm BR}(h^{\rm NP}_{95}\rightarrow b\bar b)\approx C_{h_{95}VV}^2 {\rm BR}(h^{\rm SM}_{95}\rightarrow b\bar b)\frac{\Gamma^{\rm SM}_{\rm tot,95}}{\Gamma^{\rm NP}_{\rm tot,95}}.
\end{eqnarray}

In this context, BR denotes the branching ratio, $C_{h_{95}VV}$ represents the coupling of $h_{95}$ to gauge bosons, while $C_{h_{95}uu}$, $C_{h_{95}dd}$, and $C_{h_{95}ee}$ denote the couplings of $h_{95}$ to up-type quarks, down-type quarks, and leptons, respectively. The specific forms of these couplings are as follows
\begin{eqnarray}
&&C_{h_{idd}}=\frac{-Y_d}{\sqrt{2}}{Z_H^{i1}}, ~~ C_{h_{iuu}}=\frac{-Y_u}{\sqrt{2}}{Z_H^{i2}},~~ C_{h_{iVV}}=Z_H^{i1}\cos\beta+Z_H^{i2}\sin\beta,~~ C_{h_{iee}}=C_{h_idd}.
\end{eqnarray}
Here, $Z_H$ is the unitary matrix that diagonalizes $M^2_{Total,h}$.
\section{Numerical analysis}
When analyzing the values, we take into account experimental constraints.

1. We consider the experimental constraints from the second lightest CP-even Higgs $m_{h_2}$ mass is $m_{h_2} = 125.20\pm0.11\;{\rm GeV}$\cite{PDG}.

2. $M_{Z^\prime}$ has relation with $\tan \beta_{\eta}$, which is shown in detail in Ref.\cite{tanb1}.  LHC searches limit $\tan \beta_{\eta}$ $\textless$ 1.5\cite{tanb1},
since $M_{Z^\prime}$ constraint  obtained from ATLAS and CMS analysis on several models\cite{tanb2}.

3. According to the latest LHC data\cite{limit1,limit2,limit3,limit4,limit5,limit6}, we take  the scalar lepton mass greater than 700 GeV, the chargino mass greater than 1000 GeV, and the masses of up-squarks and down-squarks greater than 1500 GeV.

4. The $Z^\prime$ boson mass is larger than 5.1 TeV. The ratio between $M_{Z^\prime}$ and its gauge $M_{Z^\prime}/g_B \geq$ 6 TeV\cite{limit7}

The values that we take into account in our numerical analysis are based on these experimental limitations.
\subsection{one-dimensional line graph}
We use images to visualize the impact of variables on the results, with the quantitative parameters set as follows
\begin{eqnarray}
&&M_{BL} = 0.2 {\rm TeV},~~ M_S = 2.7 {\rm TeV},~~ g_X = 0.3,~~ g_{YX} = -0.1,~~ \lambda_C = 0.05,~~ \nonumber\\&&T_{\lambda_C} = -0.28{\rm TeV},~~T_{\kappa} = 1.5{\rm TeV},~~l_W = 4 {\rm TeV^2},~~\xi = 17 {\rm TeV},~~ B_\mu = 1 {\rm TeV^2},\nonumber\\&&
Y_{X11} = Y_{X22} = Y_{X33} = 1,~ m_{\tilde{Q}11}^2 = m_{\tilde{Q}22}^2 = m_{\tilde{Q}33}^2 = 4.5 {\rm TeV^2},~\nonumber\\&&
 m_{\tilde{U}11}^2 = m_{\tilde{U}22}^2 =  m_{\tilde{U}33}^2 = 6.6 {\rm TeV^2},~~
m_{\tilde{D}11}^2 = m_{\tilde{D}22}^2 =  m_{\tilde{D}33}^2 = 5 {\rm TeV^2},~\nonumber\\&&
 m_{\tilde{E}11}^2 = m_{\tilde{E}22}^2 =  m_{{\tilde{E}}33}^2 = 5 {\rm TeV^2},~ m_{{\tilde{L}}11}^2 = m_{{\tilde{L}}22}^2 =  m_{{\tilde{L}}33}^2 = 1 {\rm TeV^2}.
\end{eqnarray}
Here, $M_{BL}$ is the mass of the superpartner for the gauge boson under the $U(1)_X$ group. Usually, the non-diagonal elements of the parameters are set to zero by default, unless explicitly stated otherwise.

\begin{figure}[ht]
\setlength{\unitlength}{5mm}
\centering
\includegraphics[width=2.5in]{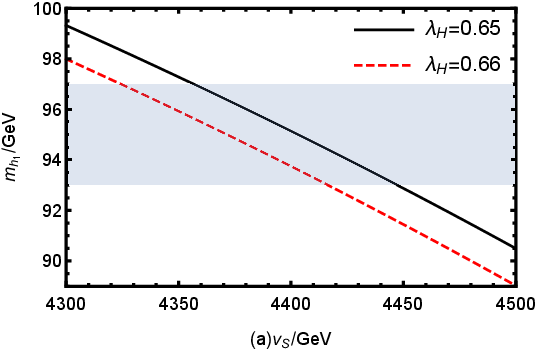}
\setlength{\unitlength}{5mm}
\centering
\includegraphics[width=2.5in]{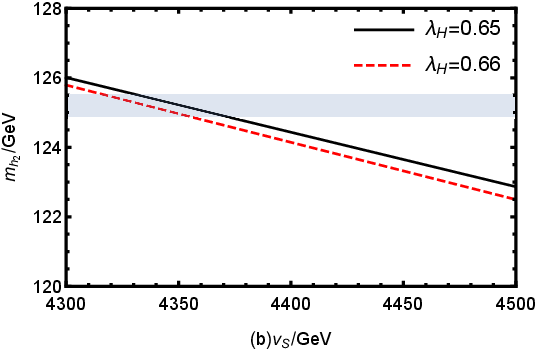}
\caption{The influences of various parameters on $m_{h_{1,2}}$: In (a) and (b), $\kappa$ = 0.15, $T_{\lambda_H}$ = 800 GeV.} {\label {xt1}}
\end{figure}
In Fig.$\ref{xt1}$(a), we plot the relationship diagram among $v_S$ and $m_{h_1}$.
In Fig.$\ref{xt1}$(b), we plot the relationship diagram between $v_S$ and $m_{h_2}$. We select the range for the mass of the lightest Higgs $(m_{h_1})$ (93 GeV $\sim$ 98 GeV), and the mass of the second-lightest Higgs $(m_{h_2})$ falls within the $3\sigma$ range (124.87 GeV $\sim$ 125.53 GeV), both of which are shaded in the diagrams. $v_S$ is the vacuum expectation value(VEV) of the added Higgs singlet $S$.
 $\lambda_H$ is the coupling constant of the term $\lambda_HSH_uH_d$ in the super potential.
The mass matrixes of neutral Higgs and several particles (neutralino, down-squark, up-squark, charged Higgs, chargino)all have the important parameters $\lambda_H$ and $v_S$, which can embody the new physics effect. In Fig.$\ref{xt1}$ (a) and (b), as $v_S$ increases, the masses of both $m_{h_1}$ and $m_{h_2}$ decrease. $v_S$, as a parameter of new physics, exhibits decoupling effects when its mass is sufficiently large, resulting in a gradual reduction of its influence. This is the primary reason why $v_S$ leads to the decrease in the masses of the two Higgs bosons. From the distribution of solid and dashed lines, when $\lambda_H$ increases slightly, the mass of $m_{h_1}$ decreases significantly, while the decrease in the mass of $m_{h_2}$ is not as noticeable. This demonstrates that $v_S$ is a sensitive parameter for both $m_{h_1}$ and $m_{h_2}$, while $\lambda_H$ is a sensitive parameter for $m_{h_1}$ but relatively insensitive and mild for $m_{h_2}$.

For the narrow interval of $v_S$, the explanation is given out. $v_S$ appears not only in Higgs tree-level mass matrix but also in loop corrections including particle mass and it represents a highly complex function that can only be analyzed graphically. When performing numerical calculations, we need to ensure that the mass of $m_{h_2}$ fall in the $3\sigma$ range and the mass of $m_{h_1}$ fall in the $93\sim98$ GeV range simultaneously, which can make some parameter intervals narrow. The value of $v_S$ is generally 3000$ \sim$6000 GeV, and due to the strict restrictions of $m_{h_2}$ and $m_{h_1}$, the range of $v_S$ becomes narrow. If $v_S$ is taken as a different value, then the other parameters will change accordingly, so we fix it in an interval for analysis. For example, in our used parameter space, the range of $v_S$ between 4300 GeV and 4500 GeV is chosen because it satisfies the mass constraints from both Higgs bosons, while other ranges of $v_S$ do not.
\begin{figure}[ht]
\includegraphics[width=2.5in]{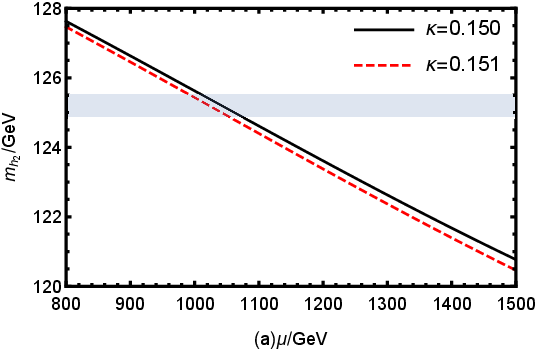}
\setlength{\unitlength}{5mm}
\centering
\includegraphics[width=2.4in]{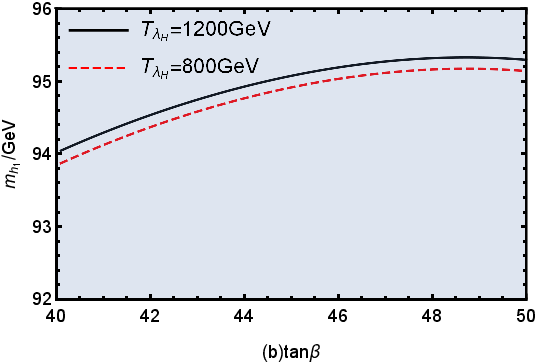}
\caption{The influences of various parameters on $m_{h_{1,2}}$:in (a), $\lambda_H$ = 0.65, $T_{\lambda_H}$ = 800 GeV, and in(b), $\lambda_H$ = 0.65, $\kappa$ = 0.15.} {\label {xt2}}
\end{figure}

$m_{h_2}$ versus $\mu$ is plotted in the  Fig.$\ref{xt2}$ (a).  $\mu$ is the Higgsino mass. As $\mu$ increases, the mass of $m_{h_2}$ gradually decreases.  When $\mu$ remains constant, as $\kappa$ increases, there is a tendency for the mass of $m_{h_2}$ to decrease, although the impact is relatively minor. Since $\kappa$ is a parameter highly sensitive to $m_{h_1}$, we select value for $\kappa$ within the range of $(0.150 \sim 0.151)$ to satisfy the condition that the mass of $m_{h_1}$ is near 95 GeV.

 In Fig.$\ref{xt2}$ (b), as $\tan\beta$ increases, the mass of $m_{h_1}$ gradually rises, but the rate of increase diminishes, which is due to the fact that as $\tan\beta$ reaches a certain value, its influence mechanism becomes increasingly weaker.
  $T_{\lambda_H}$ is the coupling constant of $T_{\lambda_H}SH_uH_d$.
    When $T_{\lambda_H}$ increases, there is a tendency for the mass of $m_{h_1}$ to increase, but the impact is relatively minor. Therefore, $T_{\lambda_H}$ is not a sensitive parameter for $m_{h_1}$.
\begin{figure}[ht]
\setlength{\unitlength}{5mm}
\centering
\includegraphics[width=2.5in]{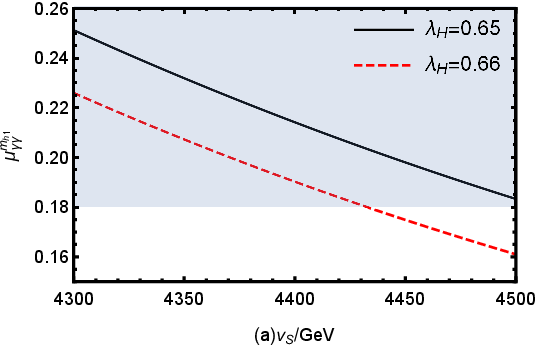}
\setlength{\unitlength}{5mm}
\centering
\includegraphics[width=2.5in]{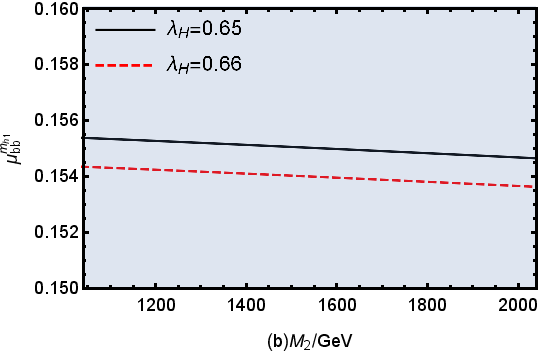}
\caption{The influence of various parameters on $\mu^{m_{h_1}}_{\gamma\gamma,b\bar b}$: In (a), $\mu$ = 1050 GeV, $\tan\beta$ = 50, $\kappa$ = 0.15, $M_2$ = 1200 GeV, and in (b), $\kappa$ = 0.15, $\tan\beta$ = 50, $v_S$ = 4400 GeV, $\mu$ = 1050 GeV.} {\label {xt3}}
\end{figure}
\pagebreak[4]

In the Fig.$\ref{xt3}$, the shaded areas represent the $1\sigma$ ranges for each the signals, respectively.  Due to certain parametric constraints (which ensure that the mass of the lightest Higgs is near 95 GeV and the mass of the second-lightest Higgs is within the $3\sigma$ range), some values within the $1\sigma$ ranges are unachievable.
In Fig.$\ref{xt3}$ (a), we plot the diagram depicting the relationship between $v_S$ and the signal $\gamma\gamma$.
As $v_S$ increases, the $\mu^{m_{h_1}}_{\gamma\gamma}$ gradually decreases. It exhibits decoupling effects when $v_S$ increases to certain values, leading to a gradual decline in the numerical results. As $\lambda_H$ increases, the $\mu^{m_{h_1}}_{\gamma\gamma}$ gradually decreases. In Figs.$\ref{xt3}$ (b), $\mu^{m_{h_1}}_{b\bar b}$ decreases slightly as $M_2$ increases. However, the effect is not significant, so $M_2$ is a moderate parameter.

\subsection{two-dimensional scatter plot}
To better study the influence of parameters, we draw some multidimensional scatter plots based on $\chi^2$ and select values that fall within  $3\sigma$ range of $\chi^2$, ensuring a more significant correlation among the numerical values. We utilize the simplified expression of $\chi^2$ as
\begin{eqnarray}
&&\chi^2 = \sum_i(\frac{\mu^{th}_i - \mu^{exp}_i}{\delta_i})^2.\label{kf}
\end{eqnarray}
In Eq.(\ref{kf}), $\mu^{th}_i$ signifies the theoretical value for the corresponding procedure derived within $U(1)_X$SSM. The experimental data is denoted as $\mu^{exp}_i$, while $\delta_i$ represents the error encompassing both statistical and systematic components.

In the $\chi^2$, we consider the mass of Higgs $m_{h_2}=$125 GeV,
and the decays  $h_1\rightarrow\gamma\gamma ~(b\bar b$) and $h_2\rightarrow\gamma\gamma$ ($b\bar b$, $ZZ$, $WW$, $\tau\bar\tau$). The specific expression of $\chi^2$ is presented as
\begin{eqnarray}
&&\chi^2 = \Big(\frac{m^{th}_{125} - m^{exp}_{125}}{\delta_{m_{h}(125)}}\Big)^2 + \Big(\frac{\mu^{th}_{b\bar b(95)} - \mu^{exp}_{b\bar b(95)}}{\delta_{{b\bar b(95)}}}\Big)^2
+ \Big(\frac{\mu^{th}_{\gamma\gamma(95)} - \mu^{exp}_{\gamma\gamma(95)}}{\delta_{{\gamma\gamma(95)}}}\Big)^2 \nonumber\\&&
~~ ~+ \Big(\frac{\mu^{th}_{\gamma\gamma(125)} - \mu^{exp}_{\gamma\gamma(125)}}{\delta_{{\gamma\gamma(125)}}}\Big)^2
+ \Big(\frac{\mu^{th}_{b\bar b(125)} - \mu^{exp}_{b\bar b(125)}}{\delta_{{b\bar b(125)}}}\Big)^2
+ \Big(\frac{\mu^{th}_{ZZ(125)} - \mu^{exp}_{ZZ(125)}}{\delta_{{ZZ(125)}}}\Big)^2\nonumber\\&&
~~ ~ + \Big(\frac{\mu^{th}_{WW(125)} - \mu^{exp}_{WW(125)}}{\delta_{{WW(125)}}}\Big)^2
+ \Big(\frac{\mu^{th}_{\tau\bar\tau(125)} - \mu^{exp}_{\tau\bar\tau(125)}}{\delta_{{\tau\bar\tau(125)}}}\Big)^2.
\end{eqnarray}

The averaged values of the experimental data are obtained from the updated PDG\cite{PDG}, $\mu^{exp(125)}_{\gamma\gamma}$ = 1.10 $\pm$ 0.06, $\mu^{exp(125)}_{ZZ}$ = 1.02 $\pm$ 0.08, $\mu^{exp(125)}_{WW}$ = 1.00 $\pm$ 0.08, $\mu^{exp(125)}_{b\bar b}$ = 0.99 $\pm$ 0.12, $\mu^{exp(125)}_{\tau\bar\tau}$ = 0.91 $\pm$ 0.09.

\begin{figure}[ht]
\setlength{\unitlength}{5mm}
\centering
\includegraphics[width=2.5in]{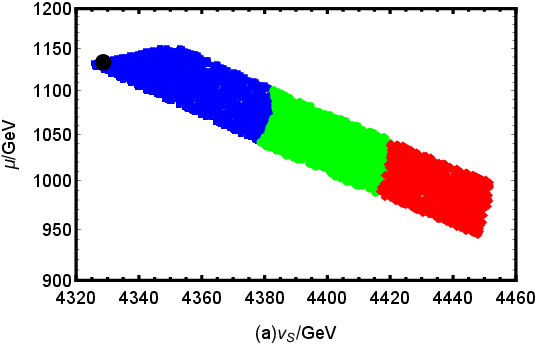}
\setlength{\unitlength}{5mm}
\centering
\includegraphics[width=2.35in]{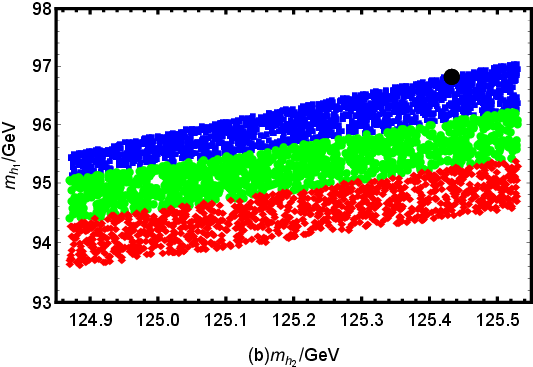}
\caption{In (a), $v_S$ versus $\mu$, and in (b), $m_{h_2}$ versus $m_{h_1}$. $\textcolor{blue}{\blacksquare}$ is a value located within the $1\sigma$ range of $\chi^2$ in the domain of $\chi^2 \leq$ 5.90. $\textcolor{green}{\bullet}$ is a value situated within the $(1\sim2)\sigma$ range of $\chi^2$, where 5.90 $\leq \chi^2 \leq$ 9.82. $\textcolor{red}{\blacklozenge}$ is a value situated within the $(2\sim3)\sigma$ range of $\chi^2$, where again 9.82 $\leq \chi^2 \leq$ 15.44. $\bullet$ represents the value corresponding to the optimal point ($\chi^2_{Best} =$ 3.62).} {\label {sd1}}
\end{figure}

\begin{figure}[ht]
\setlength{\unitlength}{5mm}
\centering
\includegraphics[width=2.5in]{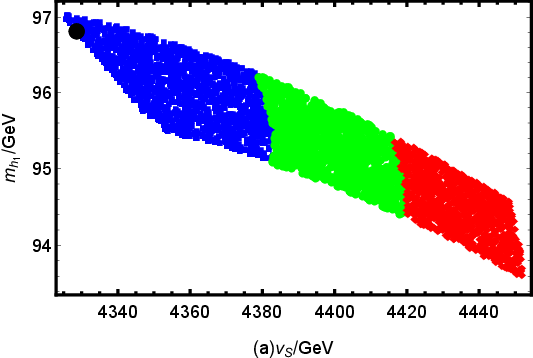}
\setlength{\unitlength}{5mm}
\centering
\includegraphics[width=2.45in]{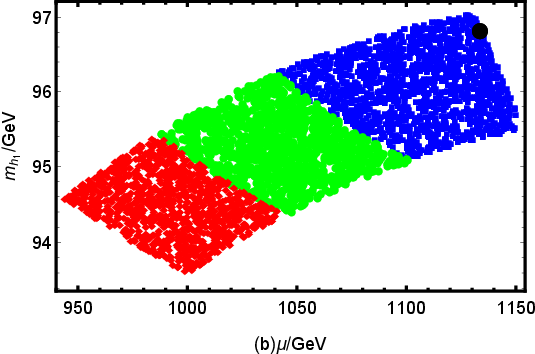}
\caption{In (a), $v_S$ versus $m_{h_1}$, and in (b), $\mu$ versus $m_{h_1}$. The marks $\textcolor{blue}{\blacksquare}$, $\textcolor{green}{\bullet}$, $\textcolor{red}{\blacklozenge}$ and $\bullet$ are same as those in the Fig.$\ref{sd1}$.} {\label {sd2}}
\end{figure}

\begin{figure}[ht]
\setlength{\unitlength}{5mm}
\centering
\includegraphics[width=2.5in]{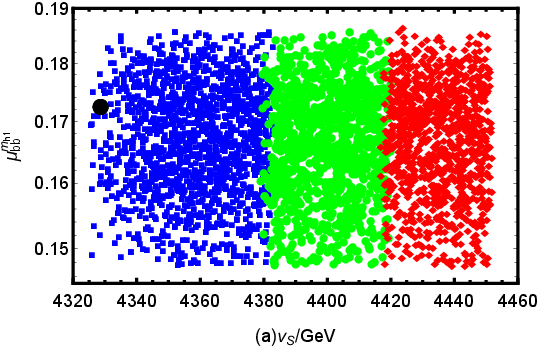}
\setlength{\unitlength}{5mm}
\centering
\includegraphics[width=2.45in]{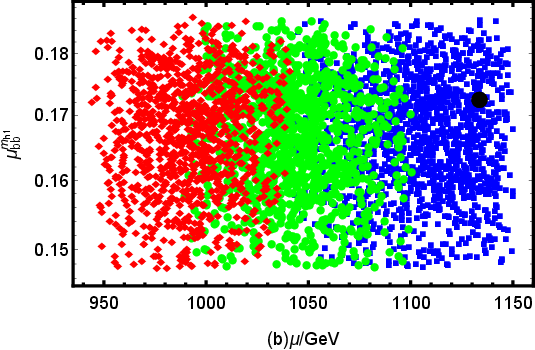}
\caption{In (a), $v_S$ versus $\mu^{m_{h_1}}_{b\bar b}$, and in (b), $\mu$ versus $\mu^{m_{h_1}}_{b\bar b}$. The marks $\textcolor{blue}{\blacksquare}$, $\textcolor{green}{\bullet}$, $\textcolor{red}{\blacklozenge}$ and $\bullet$ are same as those in the Fig.$\ref{sd1}$ and Fig.$\ref{sd2}$.} {\label {sd3}}
\end{figure}

\begin{figure}[ht]
\setlength{\unitlength}{5mm}
\centering
\includegraphics[width=2.5in]{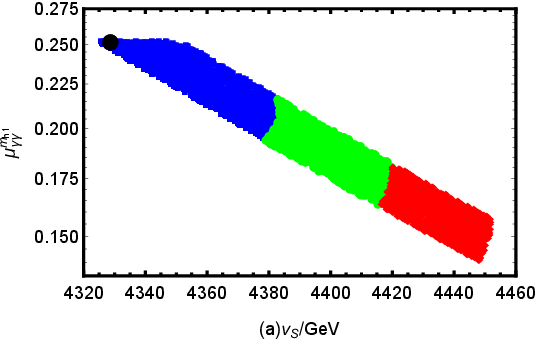}
\setlength{\unitlength}{5mm}
\centering
\includegraphics[width=2.45in]{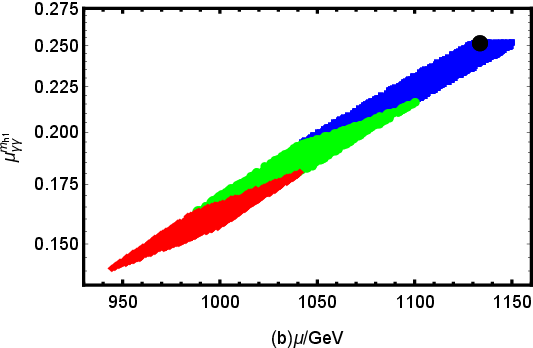}
\caption{In (a), $v_S$ versus $\mu^{m_{h_1}}_{\gamma\gamma}$, and in (b), $\mu$ versus $\mu^{m_{h_1}}_{\gamma\gamma}$. The marks $\textcolor{blue}{\blacksquare}$, $\textcolor{green}{\bullet}$, $\textcolor{red}{\blacklozenge}$ and $\bullet$ are same as those in the Fig.$\ref{sd1}$ Fig.$\ref{sd2}$ and Fig.$\ref{sd3}$.} {\label {sd4}}
\end{figure}

\begin{table*}
\caption{Scanning parameters for Fig.\ref{sd1}, Fig.\ref{sd2}, and Fig.\ref{sd3}.}
\begin{tabular*}{\textwidth}{@{\extracolsep{\fill}}|l|l|l|@{}}
\hline
Parameters&$v_S$/GeV&$\mu$/GeV\\
\hline
Min&4300~~ ~~ ~~ ~~ ~~ ~~ ~~ ~~ ~~ ~~&800~~ ~~ ~~ ~~ ~~ ~~ ~~ ~~ ~~\\
\hline
Max&4500&1500~~\\
\hline
\end{tabular*}
\label{kf1}
\end{table*}
Since we need to satisfy the masses of both $m_{h_1}$ and $m_{h_2}$, we only use two variables for plotting the scatter points(when we do the scatter plot, the mass of $m_{h_1}$ is in the range of 93 $ \sim$ 98GeV; at the same time, the mass of $m_{h_2}$ is in the range of $3\sigma$).
We randomly scan two parameters and present them in table $\textup{\ref{kf1}}$. Fig.$\ref{sd1}$, Fig.$\ref{sd2}$, Fig.$\ref{sd3}$ and Fig.$\ref{sd4}$ ($\lambda_H = 0.65, \kappa = 0.150, \tan\beta = 50$, $T_{\lambda_H} = 800$ GeV, $M_2 =$ 1200 GeV) are plotted with the parameters present in table $\textup{\ref{kf1}}$. $\textcolor{blue}{\blacksquare}$ is a value located within the $1\sigma$ range of $\chi^2$ in the domain of $\chi^2 \leq$ 5.90. $\textcolor{green}{\bullet}$ is a value situated within the $(1\sim2)\sigma$ range of $\chi^2$, where 5.90 $\leq \chi^2 \leq$ 9.82. $\textcolor{red}{\blacklozenge}$ is a value situated within the $(2\sim3)\sigma$ range of $\chi^2$, where again 9.82 $\leq \chi^2 \leq$ 15.44. $\bullet$ represents the value corresponding to the optimal point ($\chi^2_{Best} =$ 3.62).

In Fig.$\ref{sd1}$ (a), we first describe the relationship between the two variables $v_S$ and $\mu$. As $v_S$ increases, $\mu$ exhibits a downward trend, which indirectly proves that their impacts are opposite, with the optimal point located at $v_S$ $\approx$ 4330 GeV and $\mu$ $\approx$ 1130 GeV. At the optimal point, as we move outward from the center, their values are layered. In Fig.$\ref{sd1}$ (b), we plot a graph depicting the relationship between the mass of the second-lightest Higgs $m_{h_2}$ and the mass of the lightest Higgs $m_{h_1}$. As $m_{h_2}$ increases, $m_{h_1}$ tends to rise. The optimal point is located near $m_{h_2}$ $\approx$ 125.4 GeV, which aligns well with the mass of the CP-even Higgs. At this point, the mass of the lightest Higgs $m_{h_1}$ is 96.8 GeV. In the whole, the points look like a trapezium.

In Fig.$\ref{sd2}$ (a) and (b), we plot two graphs showing the relationship between $v_S$, $\mu$ and the mass of the lightest Higgs $m_{h_1}$. When $v_S$ increases, $m_{h_1}$ gradually decreases. On the other hand, when $\mu$ increases, $m_{h_1}$ gradually increases. In Fig.$\ref{sd2}$ (b), although they have a positive correlation, as $\mu$ increases, the increasing trend of $m_{h_1}$ diminishes. This is because when $\mu$ reaches a certain value, its influence gradually weakens and no longer dominates. In Fig.$\ref{sd2}$ (a) and (b), these points are uniformly distributed across the three ranges of the $\chi^2$: (0 $\sim$ 1$\sigma$, 1 $\sim$ 2$\sigma$, and 2 $\sim$ 3$\sigma$).

In Fig.$\ref{sd3}$ (a), we plot a graph showing the relationship between $v_S$ and $\mu^{m_{h_1}}_{b\bar b}$, and another graph Fig.$\ref{sd3}$ (b) depicting the relationship between $\mu$ and $\mu^{m_{h_1}}_{b\bar b}$. In Fig.$\ref{sd3}$ (a), with the increase of $v_S$, the $\chi^2$ corresponding to $\mu^{m_{h_1}}_{b\bar b}$ gradually increases. These points are neatly distributed across the three ranges of the $\chi^2$: (0 $\sim$ 1$\sigma$, 1 $\sim$ 2$\sigma$, and 2 $\sim$ 3$\sigma$) with the majority falling within the 0 $\sim$ 1$\sigma$ range. In Fig.$\ref{sd3}$ (b), as $\mu$ increases, $\chi^2$ corresponding to $\mu^{m_{h_1}}_{b\bar b}$ decreases. As $\mu$ increases, these points are uniformly distributed across the three ranges from left to right (2 $\sim$ 3$\sigma$, 1 $\sim$ 2$\sigma$, and 0 $\sim$ 1$\sigma$). The Fig.$\ref{sd3}$ (a) and Fig.$\ref{sd3}$ (b) are similar in shape, but the colors are in reverse order. Numerically, $v_S$ and $\mu$ have opposite effects on $\mu^{m_{h_1}}_{b\bar b}$.

In Fig.$\ref{sd4}$, we present the effects of $v_S$ and $\mu$ on $\mu^{m_{h_1}}_{\gamma\gamma}$ separately. In Fig.$\ref{sd4}$ (a), as $v_S$ increases, $\mu^{m_{h_1}}_{\gamma\gamma}$ gradually decreases. The maximum value is around 0.250, and the minimum value is around 0.125. These points are distributed according to the $3\sigma$ range of the $\chi^2$, with the majority of points falling within the $0 \sim 1\sigma$ range. The other points showed an almost uniform downward trend, distributed in $1 \sim 2\sigma$ and $2 \sim 3\sigma$ ranges. In Fig.$\ref{sd4}$ (b), $\mu^{m_{h_1}}_{\gamma\gamma}$ increases as $\mu$ increases, and its rate of growth is almost identical to the rate of decrease observed in Fig.$\ref{sd4}$ (a). Therefore, in the $\chi^2$, the roles of $v_S$ and $\mu$  in influencing $\mu^{m_{h_1}}_{\gamma\gamma}$ are similar to same extent. These points are arranged in an almost conical shape.
\section{Conclusion}

This work systematically investigates the 95 GeV excesses observed in the $\gamma\gamma$ and $b\bar b$ channels at the LHC within the framework of $U(1)_XSSM$. In our used parameter space, the numerical results indicate that the optimal point is located near $v_S$ $\approx$ 4330 GeV and $\mu$ $\approx$ 1130 GeV, with $\mu^{m_{h_1}}_{\gamma\gamma}$ $\approx$ 0.25, and $\mu^{m_{h_1}}_{b\bar b}$ $\approx$ 0.17, at this time.
In this condition, the lightest Higgs mass $m_{h_1}$ is 96.8 GeV, and the second-lightest Higgs mass $m_{h_2}$ is 125.4 GeV, which are consistent with the masses of CP-even Higgs bosons. These results demonstrate $U(1)_XSSM$'s capacity to concisely explain the anomalous signals through a self-consistent parameter configuration.

The second run of the LHC at $\sqrt{s}=13 ~{\rm TeV}$ places strict constraints on SUSY parameter space.
The limits at the LHC are reflected by a summary of
 results from the ATLAS\cite{ATLASSUSY} and CMS\cite{CMSSUSY} experiments.
 The gluino mass is at about 2.45 TeV. The masses of electroweak
 gauginos should be around 400-1100 GeV. Slepton mass is approximately not smaller than 700 GeV and the constraints of squark masses are also discussed. In this work, we consider the above constraints.

In our previous works \cite{WTTJHEP,WTTEPJC}, we have studied lepton flavor violating (LFV) precesses
such as $l_j\rightarrow l_i+\gamma,~ l_j\rightarrow 3l_i(i,~j=1,~2,~3,~ i\neq j), \tau\rightarrow P+l~(l=e,~\mu)$ in the $U(1)_XSSM$ and some other SUSY models.
The results imply that the non-diagonal elements of some soft breaking parameters involving the initial and final leptons are
 main sensitive parameters and LFV sources. When we take these the non-diagonal elements as zero,
 the LFV precesses are all satisfied.
 The higgs quark flavor violating decay $h\rightarrow bs$ \cite{Gao2024} is also researched in $U(1)_XSSM$, which has similar characteristic.
Moreover, in MSSM, the lightest neutralino can be dark matter candidate\cite{MSSMdark1,MSSMdark2,MSSMdark3}, and has been studied for many years.
While, in $U(1)_XSSM$ scalar neutrino can be dark matter candidate besides neutralino.
Dark matter is restricted by relic density and direct detection experiment.
We have researched the both situations for dark matter in the works\cite{U1X3,ZSMNPB21}.
Therefore, the used parameter space in this work can be coordinated to satisfy the constraints from dark matter.

This study not only advances the interpretation of the 95 GeV excesses but also highlights the $U(1)_XSSM$'s potential as a unified platform for addressing open questions in Higgs physics, FLV decays, and dark matter.

\begin{acknowledgments}
This work is supported by National Natural Science Foundation of China (NNSFC)(No.12075074),
Natural Science Foundation of Hebei Province(A2020201002, A2023201040, A2022201022, A2022201017, A2023201041),
Natural Science Foundation of Hebei Education Department (QN2022173),
Post-graduate's Innovation Fund Project of Hebei University (HBU2024SS042),
This work is supported by the Project of the China Scholarship Council (CSC) No. 202408130113. X. Dong acknowledges support from Funda\c{c}\~{a}o para a Ci\^{e}ncia e a Tecnologia (FCT, Portugal) through the projects CFTP FCT Unit UIDB/00777/2020 and UIDP/00777/2020.
\end{acknowledgments}

\appendix
\section{Parse the results}
Here, we show the examples of partical potential functions for chargino, up-type scalar quark and charged Higgs.
\begin{eqnarray}
&&\frac{\partial{V_{\tilde{\chi}^\pm}}}{\partial{\phi_d}} = -\frac{1}{16\pi^2}X_1\Big(f(Q^2,m^2_{\tilde{\chi}^{\pm}_{1}})-f(Q^2,m^2_{\tilde{\chi}^{\pm}_{2}})\Big), \nonumber\\&& \frac{\partial{V_{\tilde{\chi}^\pm}}}{\partial{\phi_{u}}} = -\frac{1}{16\pi^2}\Big(X_2f(Q^2,m^2_{\tilde{\chi}^{\pm}_{1}})+X_3f(Q^2,m^2_{\tilde{\chi}^{\pm}_{2}})\Big), \nonumber\\&&
\frac{\partial{V_{\tilde{\chi}^\pm}}}{\partial{\phi_{\eta}}} = \frac{\partial{V_{\tilde{\chi}^\pm}}}{\partial{\phi_{\bar{\eta}}}} = 0, ~~ \frac{\partial{V_{\tilde{\chi}^\pm}}}{\partial{\phi_{S}}} = -\frac{1}{16\pi^2}\Big(X_4f(Q^2,m^2_{\tilde{\chi}^{\pm}_{1}})+X_5f(Q^2,m^2_{\tilde{\chi}^{\pm}_{2}})\Big), \nonumber\\&&
\frac{\partial{V_{\tilde{u}}}}{\partial{\phi_d}} = \frac{3}{32\pi^2}\Big(X_6f(Q^2,m^2_{\tilde{u}_{1}})+X_7f(Q^2,m^2_{\tilde{u}_{2}})\Big),\nonumber\\&&
\frac{\partial{V_{\tilde{u}}}}{\phi_{u}} = \frac{3}{32\pi^2}\Big(X_8f(Q^2,m^2_{\tilde{u}_{1}})+X_9f(Q^2,m^2_{\tilde{u}_{2}})\Big),\nonumber\\&&
\frac{\partial{V_{\tilde{u}}}}{\partial{\phi_{\eta}}} = \frac{3}{32\pi^2}\Big(X_{10}f(Q^2,m^2_{\tilde{u}_{1}})+X_{11}f(Q^2,m^2_{\tilde{u}_{2}})\Big),\nonumber\\&&
\frac{\partial{V_{\tilde{u}}}}{\partial{\phi_{\bar{\eta}}}} = \frac{3}{32\pi^2}\Big(X_{12}f(Q^2,m^2_{\tilde{u}_{1}})+X_{13}f(Q^2,m^2_{\tilde{u}_{2}})\Big),\nonumber\\&&
\frac{\partial{V_{\tilde{u}}}}{\partial{\phi_{S}}} = \frac{3}{32\pi^2}X_{14}\Big(f(Q^2,m^2_{\tilde{u}_{1}})-f(Q^2,m^2_{\tilde{u}_{2}})\Big),\nonumber\\&&
\frac{\partial{V_{H^-}}}{\partial{\phi_d}} = \frac{1}{32\pi^2}\Big(X_{15}f(Q^2,m^2_{{H^-_{1}}})+X_{16}f(Q^2,m^2_{{H^-_{2}}})\Big),\nonumber\\&&
\frac{\partial{V_{H^-}}}{\phi_{u}} = \frac{1}{32\pi^2}\Big(X_{17}f(Q^2,m^2_{{H^-_{1}}})+X_{18}f(Q^2,m^2_{{H^-_{2}}})\Big),\nonumber\\&&
\frac{\partial{V_{H^-}}}{\partial{\phi_{\eta}}} = \frac{1}{32\pi^2}\Big(X_{19}f(Q^2,m^2_{{H^-_{1}}})+X_{20}f(Q^2,m^2_{{H^-_{2}}})\Big),\nonumber\\&&
\frac{\partial{V_{H^-}}}{\partial{\phi_{\bar{\eta}}}} = \frac{1}{32\pi^2}\Big(X_{21}f(Q^2,m^2_{{H^-_{1}}})+X_{22}f(Q^2,m^2_{{H^-_{2}}})\Big),\nonumber\\&&
\frac{\partial{V_{H^-}}}{\partial{\phi_{S}}} = \frac{1}{32\pi^2}\Big(X_{23}f(Q^2,m^2_{{H^-_{1}}})+X_{24}f(Q^2,m^2_{{H^-_{2}}})\Big),\nonumber\\&&
\frac{\partial{V_{u}}}{\partial{\phi_{u}}} =v_uY^2_u,~~ \frac{\partial{V_{u}}}{\partial{\phi_d}} =\frac{\partial{V_{u}}}{\partial{\phi_{\eta}}}=\frac{\partial{V_{u}}}{\partial{\phi_{\bar{\eta}}}}=\frac{\partial{V_{u}}}{\partial{\phi_{S}}}=0,\nonumber\\&&
~~ ~~ ~~\cdots,
\end{eqnarray}

where
\begin{eqnarray}
&&f(Q^2,m^2_{\tilde{\chi}^{\pm}_{1,2}}) = 2m^2_{\tilde{\chi}^\pm_{1,2}}(\log{\frac{m^2_{\tilde{\chi}^{\pm}_{1,2}}}{Q^2}}-1),~~ f(Q^2,m^2_{\tilde{u}_{1,2}}) = 2m^2_{\tilde{u}_{1,2}}(\log{\frac{m^2_{\tilde{u}_{1,2}}}{Q^2}}-1),\nonumber\\&&
f(Q^2,m^2_{u})=2m^2_{u}(\log{\frac{m^2_{u}}{Q^2}}-1),~~ f(Q^2,m^2_{H^-_{1,2}}) = 2m^2_{H^-_{1,2}}(\log{\frac{m^2_{H^-_{1,2}}}{Q^2}}-1),\nonumber\\&&
Y_u=\frac{\sqrt{2}m_u}{v_u},\nonumber\\&&
X_1 = \sqrt{2} g_2M_2\Big(4b^2+(a-c)^2 \Big)^{-\frac{1}{2}}, \nonumber\\&&
X_2 = \frac{1}{2}g^2_2v_u+\frac{1}{2}\Big(g^2_2v_u(a-c)+4b(\frac{1}{\sqrt{2}}g_2\mu+\frac{1}{2}g_2\lambda_H v_S)\Big)\Big(4b^2+(a-c)^2\Big)^{-\frac{1}{2}},\nonumber\\&&
X_3 = \frac{1}{2}g^2_2v_u-\frac{1}{2}\Big(g^2_2v_u(a-c)+4b(\frac{1}{\sqrt{2}}g_2\mu+\frac{1}{2}g_2\lambda_H v_S)\Big)\Big(4b^2+(a-c)^2\Big)^{-\frac{1}{2}},\nonumber\\&&
X_4=\frac{1}{2\sqrt{2}}\lambda_H+\frac{1}{4}\Big(-\sqrt{2}\lambda_H(a-c)+4g_2\lambda_Hv_ub\Big),\nonumber\\&&
X_5=\frac{1}{2\sqrt{2}}\lambda_H-\frac{1}{4}\Big(-\sqrt{2}\lambda_H(a-c)+4g_2\lambda_Hv_ub\Big),\nonumber\\&&
X_6=\frac{1}{8}(g^2_1g^2_{YX}+2g_{YX}g_X+g^2_X-g^2_2)v_d\nonumber\\&&
~~ ~~ +\frac{1}{8}(g^2_1g^2_{YX}+2g_{YX}g_X+g^2_X+g^2_2)(d-f)\Big(4e^2+(d-f)^2\Big)^{-\frac{1}{2}}v_d,\nonumber\\&&
X_7=\frac{1}{8}(g^2_1g^2_{YX}+2g_{YX}g_X+g^2_X-g^2_2)v_d\nonumber\\&&
~~ ~~ -\frac{1}{8}(g^2_1g^2_{YX}+2g_{YX}g_X+g^2_X+g^2_2)(d-f)\Big(4e^2+(d-f)^2\Big)^{-\frac{1}{2}}v_d,\nonumber\\&&
X_8=-\frac{1}{8}(g^2_1+2g_{YX}g_X+g^2_{YX}+g^2_2+g^2_X)\nonumber\\&&
~~ ~~ +\Big(\sqrt{2}jT_u+\frac{1}{24}(5g^2_1+8g_{YX}g_X+5g^2_{YX}-3g^2_2-3g^2_X)v_u\Big)\Big(4e^2+(d-f)^2\Big)^{-\frac{1}{2}}+v_uY^2_u,\nonumber\\&&
X_9=-\frac{1}{8}(g^2_1+2g_{YX}g_X+g^2_{YX}+g^2_2+g^2_X)\nonumber\\&&
~~ ~~ -\Big(\sqrt{2}jT_u+\frac{1}{24}(5g^2_1+8g_{YX}g_X+5g^2_{YX}-3g^2_2-3g^2_X)v_u\Big)\Big(4e^2+(d-f)^2\Big)^{-\frac{1}{2}}+v_uY^2_u,\nonumber\\&&
X_{10}=\frac{1}{4}(g_{YX}g_X+g^2_X)v_{\eta}+\frac{1}{12}\Big((5g_{YX}g_X+g^2_X)(d-f)\Big)\Big(4e^2+(d-f)^2\Big)^{-\frac{1}{2}}v_{\eta},\nonumber\\&&
X_{11}=\frac{1}{4}(g_{YX}g_X+g^2_X)v_{\eta}-\frac{1}{12}\Big((5g_{YX}g_X+g^2_X)(d-f)\Big)\Big(4e^2+(d-f)^2\Big)^{-\frac{1}{2}}v_{\eta},\nonumber\\&&
X_{12}=-\frac{1}{4}(g_{YX}g_X+g^2_X)v_{\bar\eta}+\frac{1}{12}\Big((5g_{YX}g_X+3g^2_X)(d-f)\Big(4e^2+(d-f)^2\Big)^{-\frac{1}{2}}v_{\bar\eta},\nonumber\\&&
X_{13}=-\frac{1}{4}(g_{YX}g_X+g^2_X)v_{\bar\eta}-\frac{1}{12}\Big((5g_{YX}g_X+3g^2_X)(d-f)\Big(4e^2+(d-f)^2\Big)^{-\frac{1}{2}}v_{\bar\eta},\nonumber\\&&
X_{14}=-\Big(4j^2+(d-f)^2\Big)^{-\frac{1}{2}}v_d\lambda_HY_uj,\nonumber\\&&
X_{15}=\frac{g^2_2v_d}{4}+\Big((-\lambda^2_Hv_u+\frac{g^2_2v_u}{2})k+\frac{1}{4}(g-n)(g^2_1+g^2_X+g^2_{YX}+2g_{YX}g_X)v_d\Big)\Big(4h^2+(g-n)^2\Big)^{-\frac{1}{2}},\nonumber\\&&
X_{16}=\frac{g^2_2v_d}{4}-\Big((-\lambda^2_Hv_u+\frac{g^2_2v_u}{2})k+\frac{1}{4}(g-n)(g^2_1+g^2_X+g^2_{YX}+2g_{YX}g_X)v_d\Big)\Big(4h^2+(g-n)^2\Big)^{-\frac{1}{2}},\nonumber\\&&
X_{17}=\frac{1}{4}(g^2_1+g^2_2+g^2_X+g^2_{YX}+2g_{YX}g_X)v_u+(-v_d\lambda^2_H+\frac{g^2_2v_d}{2})\Big(4k^2+(g-n)^2\Big)^{-\frac{1}{2}}k,\nonumber\\&&
X_{18}=\frac{1}{4}(g^2_1+g^2_2+g^2_X+g^2_{YX}+2g_{YX}g_X)v_u-(-v_d\lambda^2_H+\frac{g^2_2v_d}{2})\Big(4k^2+(g-n)^2\Big)^{-\frac{1}{2}}k,\nonumber\\&&
X_{19}=\frac{1}{4}g^2_Xv_\eta+\Big(\lambda_H\lambda_Cv_{\bar\eta}k+(g-n)(\frac{1}{2}g_{YX}g_X+\frac{1}{4}g^2_X)v_\eta\Big)\Big(4k^2+(g-n)^2\Big)^{-\frac{1}{2}},\nonumber\\&&
X_{20}=\frac{1}{4}g^2_Xv_\eta-\Big(\lambda_H\lambda_Cv_{\bar\eta}k+(g-n)(\frac{1}{2}g_{YX}g_X+\frac{1}{4}g^2_X)v_\eta\Big)\Big(4k^2+(g-n)^2\Big)^{-\frac{1}{2}},\nonumber\\&&
X_{21}=\frac{1}{4}g^2_Xv_{\bar\eta}+\Big(\lambda_H\lambda_Cv_{\eta}k+(g-n)(-\frac{1}{2}g_{YX}g_X-\frac{1}{4}g^2_X)v_{\bar\eta}\Big)\Big(4k^2+(g-n)^2\Big)^{-\frac{1}{2}},\nonumber\\&&
X_{22}=\frac{1}{4}g^2_Xv_{\bar\eta}-\Big(\lambda_H\lambda_Cv_{\eta}k+(g-n)(-\frac{1}{2}g_{YX}g_X-\frac{1}{4}g^2_X)v_{\bar\eta}\Big)\Big(4k^2+(g-n)^2\Big)^{-\frac{1}{2}},\nonumber\\&&
X_{23}=\sqrt{2}\mu\lambda_H+\lambda^2_Hv_S+(\sqrt{2}T_{\lambda_H}+2\sqrt{2}\lambda_HM_S)\Big(4h^2+(g-n)^2\Big)^{-\frac{1}{2}}k,\nonumber\\&&
X_{24}=\sqrt{2}\mu\lambda_H+\lambda^2_Hv_S-(\sqrt{2}T_{\lambda_H}+2\sqrt{2}\lambda_HM_S)\Big(4h^2+(g-n)^2\Big)^{-\frac{1}{2}}k,\nonumber\\&&
a=M^2_2+\frac{1}{2}g^2_2v^2_u,~~ b=\frac{1}{\sqrt{2}}g_2(v_dM_2+v_u\mu)+\frac{1}{2}g_2\lambda_Hv_uv_S,~~ c=\frac{1}{\sqrt{2}}\lambda_Hv_S+\mu,\nonumber\\&&
d=\frac{1}{24}\Big((g^2_1-3g^2_2+g^2_{YX}+g_{YX}g_X)(v^2_u-v^2_d)+2g_{YX}g_X(v^2_{\bar\eta}-v^2_\eta)\Big)+m^2_{\tilde{Q}}+\frac{v^2_uY^2_u}{2},\nonumber\\&&
j=-\frac{1}{2}\Big(\sqrt{2}(v_d\mu Y_u-v_uT_u)+v_dv_S\lambda_HY_u,\nonumber\\&&
f=\frac{1}{24}\Big((4g^2_1+4g^2_{YX}+3g^2_X+7g_{YX}g_X)(v^2_d-v^2_u)\nonumber\\&&
~~ ~~ +2(3g^2_X+4g_{YX}g_X)(v^2_{\bar\eta}-v^2_\eta)\Big)+m^2_{\tilde{U}}+\frac{v^2_uY^2_u}{2},\nonumber\\&&
g=\frac{1}{8}\Big((g^2_2+g^2_X)v^2_d+(g^2_2-g^2_X)v^2_u+(g^2_1+g^2_{YX})(v^2_d-v^2_u)+2g^2_X(v^2_\eta-v^2_{\bar\eta})\nonumber\\&&
~~ ~~ +2g_{YX}g_X(v^2_d-v^2_u+v^2_\eta-v^2_{\bar\eta})\Big)+\sqrt{2}v_S\Re{(\mu\lambda^*_H)}+\frac{v^2_S \lambda^2_H}{2}+\mu^2,\nonumber\\&&
k=\frac{1}{2}\Big(2(\lambda_Hl^*_w+B_\mu)+\lambda_H(2\sqrt{2}v_SM^*_S-v_dv_u\lambda^*_H+v_\eta v_{\bar\eta}\lambda^*_C+\sqrt{2}v_ST_{\lambda_H})\Big)+\frac{g^2_2}{4}v_dv_u,\nonumber\\&&
n=\frac{1}{8}\Big((g^2_2-g^2_X)v^2_d+(g^2_2+g^2_X)v^2_u+(g^2_1+g^2_{YX})(v^2_u-v^2_d)-2g^2_X(v^2_\eta-v^2_{\bar\eta})\nonumber\\&&
~~ ~~ +2g_{YX}g_X(v^2_u-v^2_d-v^2_\eta+v^2_{\bar\eta})\Big)+\sqrt{2}v_S\Re{(\mu\lambda^*_H)}+\frac{v^2_S \lambda^2_H}{2}+\mu^2,\nonumber\\&&
~~ ~~ ~~\cdots .
\end{eqnarray}

\end{document}